\documentclass[final,5p,times,twocolumn,authoryear]{elsarticle}

\usepackage{graphicx}%
\usepackage{multirow}%
\usepackage{amsmath,amssymb,amsfonts}%
\usepackage{amsthm}%
\usepackage{mathrsfs}%
\usepackage[title]{appendix}%
\usepackage{xcolor}%
\usepackage{textcomp}%
\usepackage{manyfoot}%
\usepackage{booktabs}%
\usepackage{algorithm}%
\usepackage{algorithmicx}%
\usepackage{algpseudocode}%
\usepackage{listings}%
\usepackage{tabularx}
\usepackage{array}
\newcolumntype{M}[1]{>{\centering\arraybackslash}m{#1}}
\usepackage{adjustbox}
\usepackage{subfigure}
\usepackage[normalem]{ulem}
\usepackage{url}

\usepackage{hyperref} 

\begin{document}

\begin{frontmatter}



\title{Agreeing and Disagreeing in Collaborative Knowledge Graph Construction: An Analysis of Wikidata} 


\author[1]{Elisavet Koutsiana}

\author[1]{Tushita Yadav}

\author[1]{Nitisha Jain}

\author[1]{Albert Meroño-Peñuela}

\author[1]{Elena Simperl}

\affiliation[1]{organization={King's College London, Informatics Department},
            addressline={Strand campus, 30 Aldwych}, 
            city={London},
            postcode={WC2B 4BG}, 
            country={UK}}

\begin{abstract}
In this work, we study disagreements in discussions around Wikidata, an online knowledge community that builds the data backend of Wikipedia. Discussions are essential in collaborative work as they can increase contributor performance and encourage the emergence of shared norms and practices. While disagreements can play a productive role in discussions, they can also lead to conflicts and controversies, which impact contributor' well-being and their motivation to engage. We want to understand if and when such phenomena arise in Wikidata, using a mix of quantitative and qualitative analyses to identify the types of topics people disagree about, the most common patterns of interaction, and roles people play when arguing for or against an issue. We find that decisions to create Wikidata properties are much faster than those to delete properties and that more than half of controversial discussions do not lead to consensus. Our analysis suggests that Wikidata is an inclusive community, considering different opinions when making decisions, and that conflict and vandalism are rare in discussions. At the same time, while one-fourth of the editors participating in controversial discussions contribute legitimate and insightful opinions about Wikidata's emerging issues, they respond with one or two posts and do not remain engaged in the discussions to reach consensus. Our work contributes to the analysis of collaborative KG construction with insights about communication and decision-making in projects, as well as with methodological directions and open datasets. 
We hope our findings will help managers and designers support community decision-making and improve discussion tools and practices. 
\end{abstract}




\end{frontmatter}



\section{Introduction}
\label{sec:introduction}

Discussions play an important role in collaborative work as they help increase motivation and enhance emerging norms and practices.
Previous studies on Computer-Supported Cooperative Work (CSCW) have emphasised the importance of disagreement in collaborative work \citep{easterbrook1993cooperation,jehn1995multimethod,franco1995anatomy,de2003task}.
Disagreements in discussions can improve performance, lead to better project practices, and offer different perspectives on a subject. However, intense disagreements can cause controversies and conflicts, which could in turn negatively influence contributors' motivation and performance, subsequently reducing engagement and leading to dropouts \citep{jehn1995multimethod,franco1995anatomy,de2003task}. 
In peer-production, self-organised communities that collaborate to produce a knowledge artefact (e.g., articles, software, or knowledge bases), disagreements can lead to higher quality results as well as the bonding of the community \citep{franco1995anatomy,arazy2011information,de2021beg,kittur2010beyond}.
More research is needed to understand how peer-production communities manage disagreements in decision-making on platforms with asynchronous discussions, particularly self-organised online knowledge communities.


In peer-production communities, construction can be explicit through community discussions or implicit through co-editing of the content. \textit{Disagreements} occur in discussions when expressing different opinions over a subject or in co-editing when objecting and reverting the work of others. A continuing disagreement can lead to a \textit{controversy} or \textit{conflict}.  
Previous studies on disagreements in online peer-production systems have focused on controversies or conflicts, and how communities argue when expressing opinions. Studies have investigated disagreements in Q\&A forums \citep{ren2019discovering}, open source software development \citep{filippova2015mudslinging,filippova2016effects,elliott2002communicating}, and knowledge projects \citep{wang2016piece,Borra2015,zielinski2018computing,kittur2007he,vuong2008ranking,sumi2011edit,jankowski2014predicting,yasseri2014most}.
Identifying controversies or conflicts is essential for their resolution; however, it is important to understand the manner in which the community manages disagreement and takes decisions \citep{Borra2015,vuong2008ranking}, and the way of cooperation among the community members \citep{zielinski2018computing}. This understanding can ensure high quality content, balanced viewpoints, and prevent misinformation \citep{zielinski2018computing}. 
In addition, argument detection is essential to help participants handle the vast amount of information in a discussion and include all opinions in a decision \citep{wang2020argulens}, and get an understanding of positions \citep{sepehri2011towards}.

In this paper, we perform a comprehensive analysis on disagreements in Wikidata \citep{vrandevcic2014wikidata}, the world's largest open-source collaborative knowledge graph. \textit{Knowledge graphs} (KGs) are collections of entities and their relations structured as graphs describing a specific domain. \textit{Collaborative KGs} are created by humans in an enterprise environment or using crowd-sourcing or co-editing platforms \citep{hogan2021knowledge}. With a large number of entities (more than $100M$) and a community of approximately $24K$ active members,\footnote{\url{https://www.wikidata.org/wiki/Wikidata:Statistics}} Wikidata serves as the backend of the online encyclopedia Wikipedia and supports many intelligent systems like search engines (e.g., Google) and virtual assistance (e.g., Amazon's Alex) \citep{haase2017alexa,fensel2020knowledge,beloucif2023using}.

Our analysis focuses on the following research questions: 
\begin{itemize}
    \item $RQ1$ - Where do we find disagreements in Wikidata discussions?
    \item $RQ2$ - What are the main issues in controversial discussions in Wikidata?
    \item $RQ3$ - What are the characteristics of controversial discussions in Wikidata?
    \item $RQ4$ - How do editors argue when disagreeing in Wikidata?
    \item $RQ5$ - What are the roles of editors in argumentation in Wikidata?
\end{itemize}

To answer these research questions, we analyse the discussions recorded as textual conversations between Wikidata editors through a mixed-methods approach, including: (i) descriptive statistical analysis on the whole corpus of discussions in Wikidata; (ii) thematic analysis on a sample of discussions with disagreements, where we identify controversies; (iii) measurements of the radial tree structure of discussions \citep{herke2009radial} as well as statistical tests for the sample of discussions used in the thematic analysis; and (iv) content analysis for the discussions identified as controversial in the thematic analysis. 

This work contributes to the analysis of collaborative KG construction with: (i) insights about decision-making in Wikidata; (ii) an overview of the complete corpus of Wikidata discussions; (iii) a framework to identify discussions with intense disagreements; (iv) two coding schemes for content analysis to identify argumentation patterns and the role of participants in discussions; (v) a dataset including discussions for properties for deletion, property proposal, requests for comment channels; and (vi) annotated datasets for controversial discussions and argumentation. Our findings provide an understanding of how the Wikidata community discusses and disagrees, and offers advice to Wikidata designers and community managers to support decision-making and communication. Our study shows that:
\begin{itemize}
\item Compared to Wikipedia, there is little conflict in Wikidata ($7\%$ of the codes), and a few disruptions or vandals in discussions ($1\%$ of participants in controversies). However, despite the lack of intense conflicts, over half of the controversial discussions are closed without consensus ($62\%$). This can indicate that controversial discussions (including many participants and long content) make it difficult for the community to reach a consensus, and they need extra support.  
\item The most frequent controversial issue is Wikidata processes, i.e., disagreement on how to perform a task ($52\%$ of controversial discussions), with most arguments using examples of similar cases, policies and practices as counter examples ($33\%$ of arguments). This means that participants in argumentation need to be very well informed about policies to participate and engage in discussions. A good understanding of policies and values is essential in other peer-production projects like Wikipedia \citep{schneider2013arguments}. This suggests that managers could support the community with clear and frequently updated policies and guidelines in easy to access areas of the projects. This can enhance cooperation and decision-making. 
\item While the majority of participants in controversies ($47\%$) engage in the discussion and aim to influence others with their arguments, a high number of participants ($25\%$) do not engage in the discussion, despite the legitimate and insightful arguments. The lack of engagement combined with the lack of consensus may suggest that controversial discussions require attention, flagging the need for new tools to support when needed, control discussions, further instructions, or practice changes.
\end{itemize}

\section{Background: The Wikidata collaborative knowledge graph}
\label{sec:background}

\begin{figure*}[h]
\centering
\includegraphics[width=0.9\linewidth]{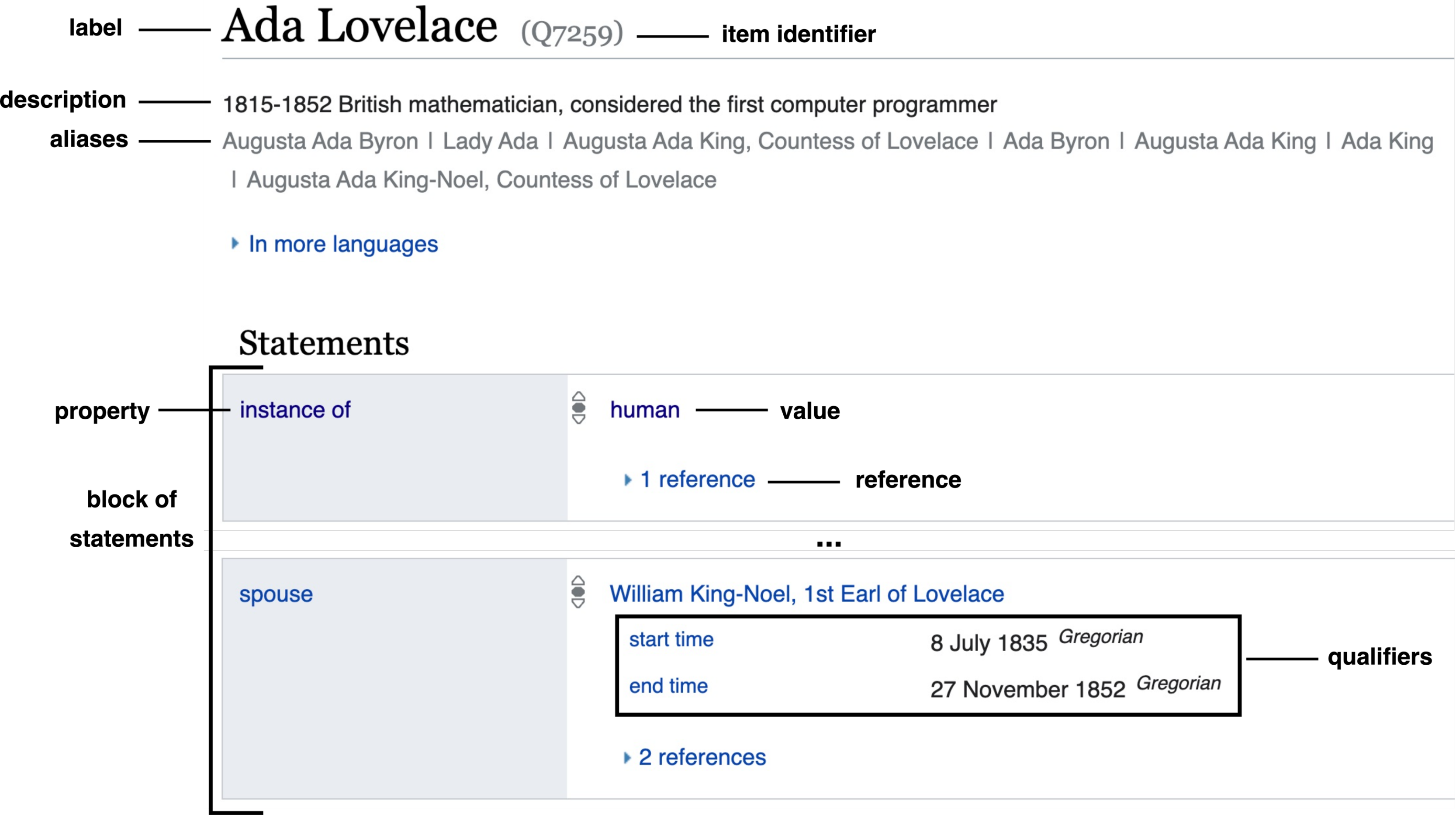}
\caption{An example of the Wikidata KG for the item \textit{Ada Lovelace}. We use red annotations to highlight the main features.}
\label{fig:wikidata}
\end{figure*}

\subsection{The Wikidata knowledge graph}
Wikidata is an online peer-production system \citep{benkler2015peer} where a community of volunteers from all over the world create and maintain a large KG \citep{vrandevcic2014wikidata}. A KG can be thought of as a technology to structure and organise factual data. In this graph, the \textit{nodes} represent the entities of interest in a domain, such as people or places. The \textit{edges} describe the entities in the form of their attributes, such as the date of birth of a person, or in the form of relations to other entities, such as the relation between a person and the place they were born in. 
Wikidata KG currently adds more than $100M$ nodes (called ``items") and $10K$ edges (called ``properties"). A glossary for Wikidata terminology can be found here \url{https://www.wikidata.org/wiki/Wikidata:Glossary}.

Furthermore, Figure~\ref{fig:wikidata} presents a node and some of its edges as an example.
The figure shows statements about the item \textit{Ada Lovelace} in the form of tuples consisting of a subject (the item that is discussed, in this case \textit{Ada Lovelace}), a property (for instance, \textit{instance of}) and the value of that property (\textit{human}). Some statements have more than three fields called qualifiers (for instance \textit{start time} and \textit{end time}) to add additional information to the item. 
Items and properties are accompanied by human-readable \textit{labels, descriptions}, and \textit{aliases} in $358$ languages \citep{vandenbussche2017linked}.\footnote{\url{https://www.wikidata.org/wiki/Wikidata:Introduction}}
Both items and properties in Wikidata are organised in taxonomies, i.e. classes. Wikidata assigns classes for instances with common features using the property \textit{`instance of'}. In the figure, the item \textit{Ada Lovelace} is an \textit{instance of} the class \textit{human}.

\subsection{Wikidata community and their contributions}
\label{sec:community and contributions}
People contribute to Wikidata in multiple ways. They can add, remove, or alter content, at various levels of granularity, ask or answer questions, discuss and decide upon different options to structure a certain piece of information, or help others do so.
Contributors are referred to as \textit{editors}, which can be either \textit{humans} or \textit{bots} (i.e., software executing simple, repetitive edits). Human editors may choose to register themselves or contribute anonymously. All editors are allowed to work on items, but registered editors have additional editing rights, depending on their experience with the project, the extent of their contributions, and the duration of their membership.\footnote{\url{https://www.wikidata.org/wiki/Wikidata:User_access_levels}} For instance, one group of editors with higher administrative editing rights are the \textit{Administrators}. They are elected by a community vote and are responsible for tasks such as deleting items, blocking editors or unauthorised bots, and protecting pages to prevent vandalism. They can also grant higher rights to other editors. An example of editors with higher technical editing rights are the \textit{Property creators}, who respond to community requests to create new properties, initiate community discussions about the request, and ensure that a decision is made.

\begin{figure*}[h]
\centering
\includegraphics[width=0.9\linewidth]{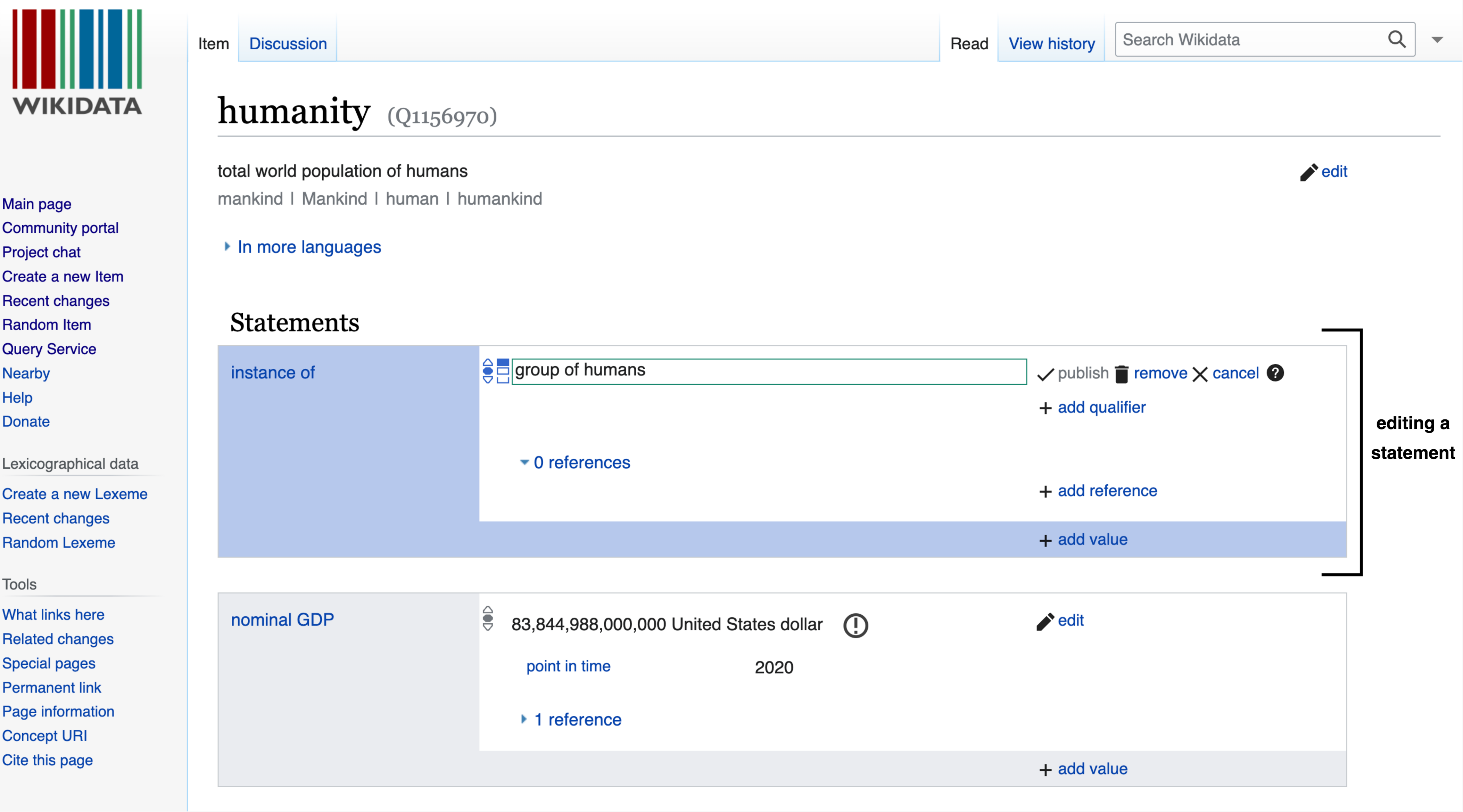}
\caption{An example of editing a statement for the Wikidata item {\fontfamily{qcr}\selectfont humanity}}
\label{fig:wikidata_editing}
\end{figure*}

\begin{figure*}[h!]
\centering
\includegraphics[width=0.9\linewidth]{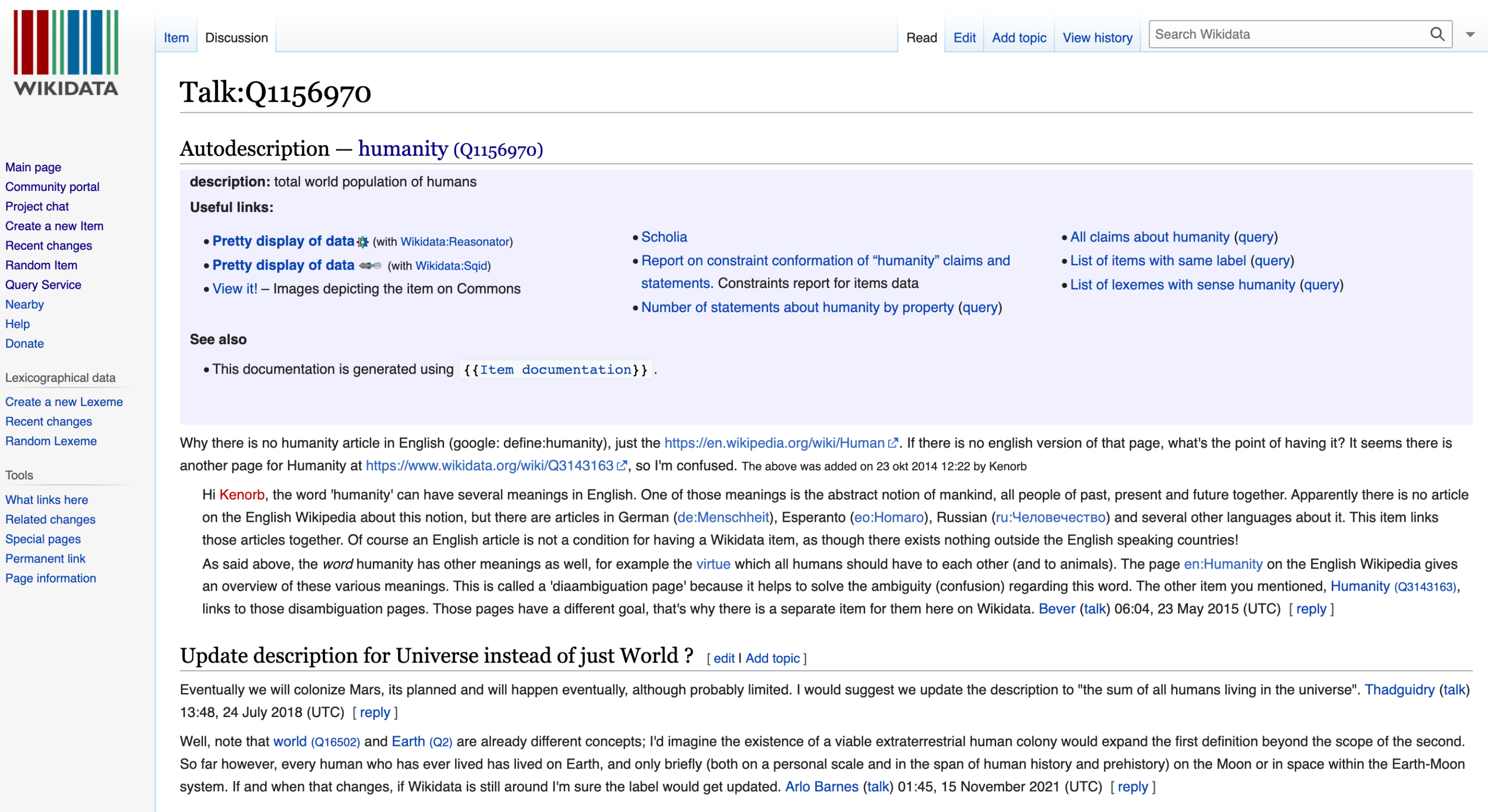}
\caption{An example of a talk page for the Wikidata item {\fontfamily{qcr}\selectfont humanity}}
\vspace{0.1in} 
\label{fig:wikidata_talk_page}
\end{figure*}

Wikidata has approximately $24K$ active registered editors, $20K$ anonymous editors, and $100$ active bots every month.\footnote{\url{https://stats.wikimedia.org/\#/metrics/wikidata.org}} 
However, for these groups, different participation patterns have been observed over the years.
According to the statistics related to the editing activities of Wikidata, published by Wikimedia Foundation (i.e., the non-profit organization hosting Wikidata), bots are responsible for $52\%$, while human editors perform $48\%$ of all edits. Previous studies on participation patterns show that almost $60\%$ of the editing activities by bots are related to item statements, i.e., adding context within a domain of interest, while $30\%$ are related to item descriptions, labels and aliases \citep{muller2015peer}. Similarly, for human editors, $13\%$ are related to item descriptions and almost $30\%$ of item statements. In addition, $30\%$ of editing activities by humans contribute to the core of the graph by describing and creating new properties and their talk pages. However, these activities come from $2\%$ of human editors \citep{muller2015peer}. Regarding human editors, it is interesting to note that only a small percentage of editing activities, $0.5\%$, comes from anonymous editors \citep{piscopo2018models}. For the registered editors, only $2\%$ of editors are responsible for $95\%$ of the edits \citep{piscopo2017wikidatians}.

From a technical point of view, Wikidata is built using a \textit{wiki}, a collaborative web platform that allows users to add and edit items and properties using web pages in a browser \citep{wagner2004wiki}. Each item or property has its own wiki page, which can be changed on the fly. Each wiki page has a discussion page, referred to as a \textit{talk} page, which is discussed in greater detail in Section~\ref{sec:discussions_WD}. Figure~\ref{fig:wikidata_editing} shows an example of how editors can edit a statement in the page of the item {\fontfamily{qcr}\selectfont humanity}, and Figure~\ref{fig:wikidata_talk_page} shows the corresponding talk page.

Wikimedia Foundation provides public access to comprehensive logs, including the history of edits for all content and discussion pages.\footnote{\url{https://www.wikidata.org/wiki/Wikidata:Data_access}}
The editing activities for each page, referred to as \textit{revisions}, include metadata such as the page name, the name of the editor responsible for the revision, a timestamp, a comment, and a text description. The comment classifies the edit into a pre-defined set of actions, such as addition, deletion, revert, undo, restore, and so on. The textual description adds context to the edit action. The data can be used for various research purposes to help and improve Wikidata practices.

\begin{table*}[h]
\caption{Lists of the discussion channels in Wikidata and their descriptions (\href{https://www.wikidata.org/wiki/Wikidata:Community_portal}{source}).}
\scriptsize
\label{tab:list_of_discussion_channels}
\centering

\begin{tabular}{|m{0.18\linewidth}|p{0.75\linewidth}|}
\hline
\textbf{Discussion Channel}       & \textbf{Description}                                             \\\hline
\textbf{Talk pages} \\\hline
Items                     & Each item page can have a corresponding page for discussions related to their construction.                           \\\hline
Properties              & Each property page can have a corresponding page for discussions related to their construction.                           \\\hline
Users                   & Each editor can have a corresponding page for discussions related to their editing activities.                           \\\hline
Wikiprojects   & Wikiprojects are groups of editors collaborating to improve Wikidata on particular topics (e.g, sports or biology). Each Wikiproject has a corresponding page for discussions to coordinate the editing activities relates to the specific topic \\\hline
Others  & Every content page (e.g. Wikidata main page, help pages) can have a corresponding page for discussions related to their content. \\\hline
\textbf{Communication pages} \\\hline
Project chat                      & General questions about how to do or what are the common practices for tasks in the project, e.g, \textit{`` How we should represent external identifiers where there is both  a permanent ID and a human-friendly ID?''}.                             \\\hline
Requests for comment             & Requests for changes in practices or guidelines, e.g, \textit{``Gender neutral labels for occupations and positions in French''}.                          \\\hline
Report a technical problem       & Discussion about platform complication, e.g, \textit{``User contributions search is borken''}.                           \\\hline
Request a query                   & Requests for Wikidata SPARQL queries, e.g, \textit{``Query for items that are both subclass/instance of something and it's opposite''}.                             \\\hline
Interwiki conflicts               & Report problems with content on other wikis, e.g, when two or more items exist that seemingly correspond to the same Wikipedia article or other Wikimedia site page.                      \\\hline
Bot requests                      & Requests for tasks to be done by a bot, e.g, \textit{``request to add identifiers from FB''}.                           \\\hline
Property proposal        & Propose the creation of a property, e.g, a suggestion to create the property ``number of penalty kicks scored''  because \textit{``Differentiate regular goals from penalty kicks, people use twice the quantifier in last editions of World Cup, rendering strange result''}                             \\\hline
Administrators' noticeboard       & Reporting vandalism, requesting page protections, etc, e.g, \textit{``Would like to draw attention to the edits of user Turktimex3''}.           \\\hline
Translators' noticeboard          & Report a translation problem, ask to mark a page for translation, e.g, \textit{``Please, mark for translation Template:AdminsChart/Title and Template:AdminsChart/Elected''}. \\\hline
Bureaucrats' noticeboard          & Requesting for flood flag, etc, e.g. editors should use the flood flag to avoid confusion when they do repetitive edits.                                  \\\hline
Requests for deletions            & Deletion requests of items and pages, e.g, in case of duplication or not following Wikidata policies like notability.                             \\\hline
Properties for deletion           & Deletion requests of properties, e.g, \textit{``P9099 (P9099) seems to be redundant to Property:P7900''}                                   \\\hline
Requests for permission & Permissions requests for permission to create a bot ot to get higher rights for editors, e.g, \textit{`` request for administrator: Trusted user who previously was overwhelmingly made a administrator. Rights were taken away due to inactivity but looking at xtools they are much more active this year''}.  \\\hline       
\end{tabular}
\end{table*}

\begin{figure*}[h]
\centering
\includegraphics[width=\linewidth]{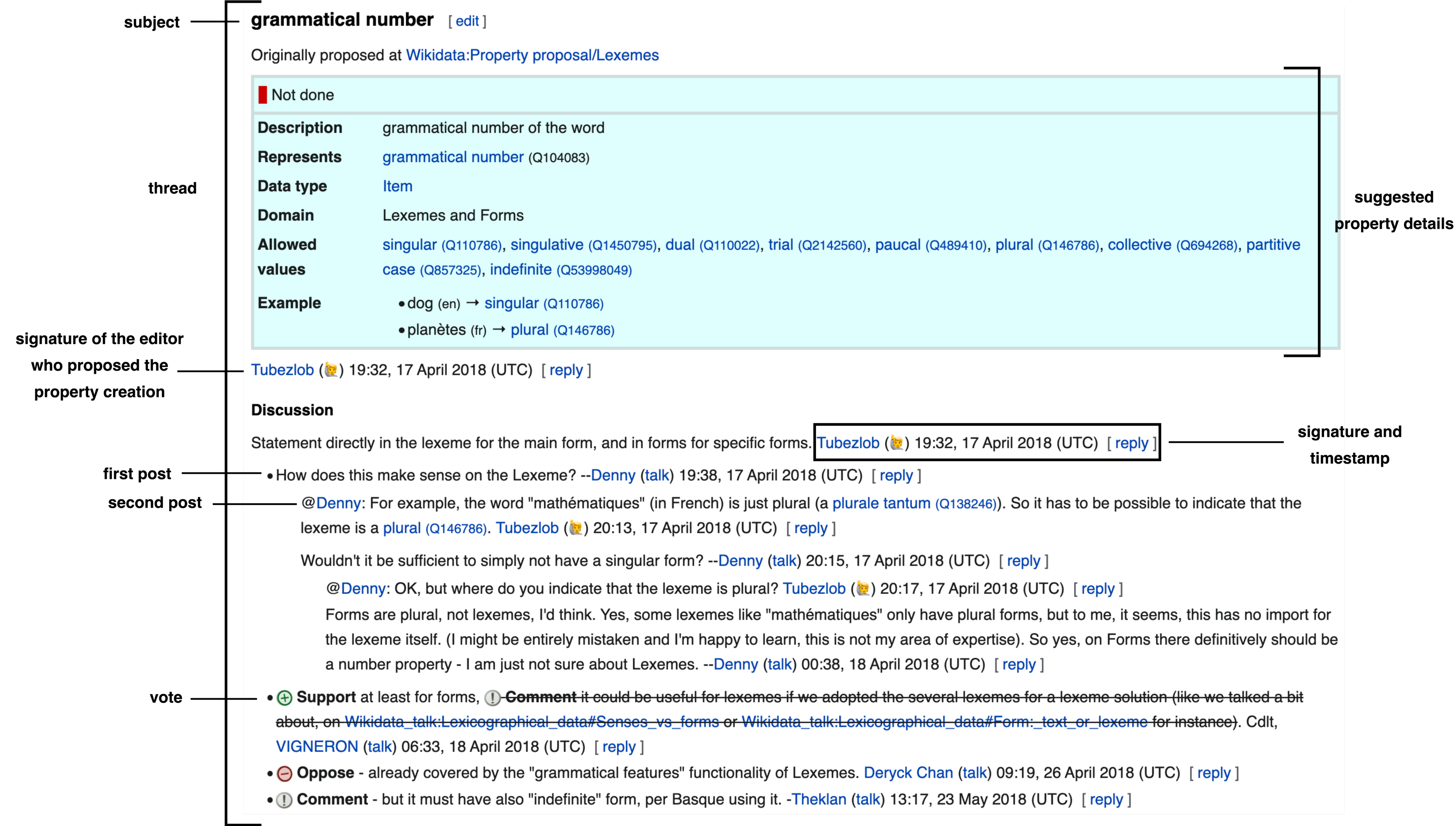}
\caption{An example of a Wikidata thread from the \textit{Property proposal} discussion channel. We use annotations to highlight the thread's main features.}
\label{fig:pp_thread}
\end{figure*}

\subsection{Discussions in Wikidata}
\label{sec:discussions_WD}

Similar to other peer-production systems, including those creating knowledge bases and KGs \citep{simperl2014collaborative}, discussions are a primary tool for collaboration in Wikidata.
Wikidata has two types of \textit{discussion channels}: the talk pages,\footnote{\url{https://www.mediawiki.org/wiki/Help:Talk_pages}} and the \textit{communication} pages.\footnote{\url{https://www.wikidata.org/wiki/Wikidata:Community_portal}} Table \ref{tab:list_of_discussion_channels} contains lists of the discussion channels and their functionality. As noted earlier, each item and property has its own talk page for the purpose of raising questions, facilitating coordination between editors, and flagging any errors.

In addition, Wikidata has various communication pages to support discussions on general concerns that are not specific to an item or property. Broadly, some communication pages are about technical support, such as with the formulation of a query to retrieve certain data from the graph, or for automating a task; data structure and organisation, such as introducing a new type of property on the graph, deleting content from the graph; as well as general and administrative concerns around user management, suspected vandalism, translations of pages, and so on. 

Editors use the discussion channels to get information, flag errors, introduce changes, suggest new directions, and request actions. Often, editors are required to make consensus and make decisions on how to proceed with the raised issues. Decision-making in Wikidata is a community-driven process that relies on consensus, discussions, guidelines, and administrative roles rather than a centralised authority. Since Wikidata is a collaboratively edited knowledge base, contributors work together to determine how data should be structured, corrected, and maintained. Our observations indicate that discussions on talk pages typically involve a limited number of participants, with decisions often reached between two to four individuals debating a topic related to the content page. In talk pages, it is uncommon for more than ten people to engage in discussions over the same issue. In some instances, a consensus is not reached, resulting in no action being taken. Additionally, on communication pages, participation in discussions tends to be higher, and a voting system is frequently employed to facilitate decision-making. Editors have the option to cast votes indicating either ``support'' or ``oppose'', providing justification for their decisions through personal opinions or by citing previous arguments presented by fellow editors. Nonetheless, it is not uncommon for editors to submit their votes without accompanying justification.
When a majority vote leads to consensus, an editor with high editing rights, who was not involved in the discussion is authorized to close it. However, there are cases where consensus is not achieved, and the discussion may remain inactive for extended periods. In such cases, discussions may either be left open for future consideration or, once again, an uninvolved editor may close the discussion without reaching a consensus, leaving the issue unresolved. It remains unclear how an editor determines whether to keep a discussion open or closed, with or without consensus, but if there are objections to the closure, those can be articulated and continue to be debated.

In our analysis, we will focus primarily on three specific communication pages, \textit{Properties for deletion, Property proposal} and \textit{Requests for comment} as we have identified these pages to be associated with the most disagreements. Section~\ref{sec:method-RQ1} presents the rationale behind the selection of these discussion channels. The \textit{Properties for deletion} and \textit{Property proposal} pages are concerned with the properties (i.e., edges) of the graph. In the former, editors suggest the deletion of properties that are not considered useful anymore or that are leading to confusion, while in the latter, editors request the creation of properties in order to complete semantic gaps and extend the graph. The \textit{Requests for comment} is a general purpose page that is concerned with a wide range of requests about changes in policies, regulations or practices, such as gender for labels in French, changes in the way editors write in the \textit{Request for comment} page.

All discussions on Wikidata follow the same format. Figure~\ref{fig:pp_thread} presents an example of discussion from the channel \textit{Property proposal}. We refer to each discussion as \textit{thread}. A thread is a set of \textit{posts} under a \textit{subject} title. At the end of each post, the name of the corresponding editor and a timestamp should be included.

Certain communication pages, require the use of a template to facilitate the discussion of new issues. This could be due to the volume of issues or to help the flow of the conversation.
The template includes a rationale for the proposal and examples of use. Once an issue is raised, it is discussed by the community. Some communication pages, like the \textit{Properties for deletion} and \textit{Requests for comment} and the requests for comment, have a limited period to be discussed, $7$ and $30$ days, respectively. 
Participants in the discussion have the opportunity to vote in support of or against a suggestion. Once a consensus is achieved, or if the thread has remained inactive for a certain period, an editor not involved in the discussion is empowered to close the request. An example of a closed discussion is presented later in Section~\ref{sec:RQ2} (Figure~\ref{fig:find_a_grave}). On top of the discussion, the status of the discussion is stated.


\section{Related work}
\label{sec:related_work}

In this section, we provide an overview of prior work that has researched disagreement and helped in shaping the research questions of this study.
Previous studies have investigated subgroups of opinions \citep{hassan2012detecting}, power relationships \citep{danescu2012echoes,biran2012detecting}, interactions \citep{brandes2009network,garimella2018quantifying,leskovec2008planetary}, and the relevances of participants \citep{alsinet2020measuring}.
A big part of the literature in peer-production focused on the study of disagreements, investigating (i) controversies and conflicts, and (ii) argumentation.

\textbf{Controversies and conflicts.} The Oxford dictionary defines controversy as a `public discussion and argument about something that many people strongly disagree about, think is bad, or are shocked by'.\footnote{\url{https://www.oxfordlearnersdictionaries.com/definition/english/controversy?q=controversy}} Studying scientific controversies, McMullin states that controversy is a `publicly and persistently maintained dispute' making the argument that `to count as controversy a disagreement must be a continuing one' \citep{mcmullin1987scientific}. In addition, according to the Oxford dictionary, a conflict is `a situation in which people, groups or countries disagree strongly or are involved in a serious argument'.\footnote{\url{https://www.oxfordlearnersdictionaries.com/definition/english/conflict_1?q=conflict}}
We find controversies and conflicts on the web, such as in debate forums \citep{beck2018managing}, social media \citep{garimella2018quantifying}, knowledge sharing forums \citep{hara2016co}, and collaborative settings \citep{filippova2015mudslinging,elliott2002communicating,wang2016piece,zielinski2018computing,kittur2007he,yasseri2014most}. In online collaboration, which is our object of study, researchers investigated the detection of controversies and conflict \citep{wang2016piece,Borra2015,zielinski2018computing,kittur2007he,ren2019discovering}, and their characteristics \citep{suh2007us,beck2018managing,filippova2015mudslinging,filippova2016effects,elliott2002communicating}.

Plenty of previous works have focused on Wikipedia for
the detection of controversial or conflictive articles \citep{zielinski2018computing,vuong2008ranking,sumi2011edit,jankowski2014predicting,yasseri2014most}, the content of the articles themselves \citep{bykau2015fine,kittur2007he,jankowski2014predicting}, or discussions related to edits on the articles \citep{wang2016piece,ho2016french}.
These studies have investigated controversies using the history of edits i.e., the sequence of revisions during their write-up (e.g., number of revisions in the article, content changed by editors, article length, article age) \citep{viegas2004studying,vuong2008ranking,sepehri2012leveraging,yasseri2014most}. Others combined the history of edits with the discussions taking place in parallel (e.g. number of revisions in the discussion page, number of editors, number of anonymous or administrative edits on the article or discussion page) \citep{kittur2007he,jankowski2014predicting}. 
To detect controversies in Wikipedia, researchers have used various methods, such as support vector machine \citep {kittur2007he,wang2016piece}, sentiment analysis \citep{wang2016piece,jankowski2014predicting} and network analysis based on edit reverts \citep{vuong2008ranking}.
Their results showed that patterns of controversy and conflict were too complex to be captured by one single characteristic, and combinations of various factors such as number and type of edits, discussion length and content must be considered \citep{rad2012identifying}. 
However, other works have shown that some features were more valuable than others for identifying controversial characteristics. \cite{kittur2007he}
proved that the most important feature in detecting conflicts was the number of revisions in the talk pages, implying that a higher number of posts on the talk pages increased the probability of conflict.
In addition, \cite{wang2016piece} used sentence level sentiment analysis and showed that their best models combined the article's topic, discussion length, number of participants, and positive/negative sentiment ranks, rather than just sentiment analysis scores. 
This was also suggested by \cite{jankowski2014predicting}, who argued that sentiment analysis alone could not predict a controversial article but could indicate its controversial parts.

In other peer-production projects such as Q\&A forums, where members discuss, ask, and answer questions related to a variety of topics such as Quora\footnote{\url{https://www.quora.com/}} and Reddit\footnote{\url{https://www.reddit.com/}} or specialised topics like Stack Overflow,\footnote{\url{https://stackoverflow.com/}} (which is for software developers), studies also aimed to detect and explore controversies in order to support community operation.
Prior work on StackOverflow \citep{ren2019discovering} created a tool to detect and summarise controversies. The authors aimed to reduce the negative impact of controversial answers to developers, and helped them find controversial discussions and take decisions.
Their experiments showed that when members used the summarised controversies tool to solve a task they had a high rate of correct answers. In addition, previous work on the Reddit community \citep{zhang2017characterizing} aimed to categorise posts in discourse act such as question, answer, and disagreement. \cite{zhang2017characterizing} found that disagreements lasted from $1$ to $7$ posts and that often ($60\%$ of the cases) did not receive answers. The study also found that the topics with the highest proportion of disagreements were ``change my view'' and ``political discussion''. However, a study by \cite{jasser2022controversial} on the Reddit forum showed that discussions staged as controversial were more likely to attract higher activity.  
While many studies in online communities focused on how to help members perform better, there were also studies interested in what made members to participate in discussions. In the general interest forum  Quora, previous work studied what influenced members to follow and discuss with other members \citep{nwadiugwu2021influencing}. \cite{nwadiugwu2021influencing} showed that members use controversies as a means to attract followers. Quora members explained that they use controversial posts to attract others' attention and participate in discussions. 

In addition, previous studies about conflicts in peer-production projects like open source software development \citep{filippova2016effects} and Wikipedia \citep{arazy2013stay} categorised conflicts into \textit{task}, \textit{affective}, \textit{process} and \textit{normative}. Task conflicts had a cognitive dimension related to disagreements on the content to be added. Affective conflicts, by contrast, involved emotional disagreement. Process conflicts occurred when people disagree on how to perform a task rather than the task itself. Finally, normative conflicts involved group function disagreements. \cite{filippova2015mudslinging,filippova2016effects} studied the impact of conflict categories in open source software development. Normative conflicts were positively correlated with dropouts; while task conflicts did not significantly affect editors' performance \citep{filippova2016effects}. Furthermore, participation in discussions about the core tasks of the project, where editors needed to take decisions, reduced normative and process conflicts \citep{filippova2016effects}. In extensive discussions over issues, it was possible for different types of conflicts to transform into others. A common case was when task conflicts transformed into affective or process conflicts \citep{filippova2015mudslinging}. 
Prior work in peer-production had also studied how conflicts arose and got resolved \citep{elliott2002communicating}. \cite{elliott2002communicating} showed that archives of past editing and discussion actions helped editors to resolve their disagreements by giving examples. Furthermore, occasional editors, people who contribute for short periods, were able to start, but also mitigate conflicts.

\textbf{Argumentation.} Looking again in the Oxford dictionary, an argument is `a conversation or discussion in which two or more people disagree, often angrily', and argumentation consists of `logical arguments used to support a theory, an action or an idea'.\footnote{\url{https://www.oxfordlearnersdictionaries.com/definition/english/argument?q=argument}} \footnote{\url{https://www.oxfordlearnersdictionaries.com/definition/english/argumentation?q=argumentation}} 
Argumentation mining is the study of automatically identifying arguments \citep{lippi2016argumentation}. \cite{schneider2014automated} emphasised the importance of argumentation mining, suggesting that its automation in debates could give personalised feedback to post convincing arguments, helped in identifying the weaknesses in each others' arguments, and offered a summary of the arguments in a debate. Furthermore, \cite{wang2020argulens} stated that argument detection in peer-production helped in handling the vast amount of opinions and avoiding overlooking arguments in the middle of the conversation.

Previous studies have analysed argumentation in online debates from different perspectives, with quantitative and qualitative means, in terms of classifying the arguments, identifying participants roles, and ranking their persuasive power. Prior work, to classify arguments used Walton's \citep{walton2008argumentation} argumentation on social media \citep{schneider2014did} and Wikipedia \citep{schneider2013arguments} with codes like \textit{argument from example, practical reasoning, argument from rules}, and so on. 
Other studies in Wikipedia, focused on knowledge sharing in arguments and looked, for example, \textit{background knowledge, facts about the topic, citation}, and so on \citep{hara2016co}, or used a dialogue scheme, such as \textit{inform, manage, evaluate}, and so on \citep{freard2010role}. 
Furthermore, similar studies in open source software development projects \citep{wang2020argulens} and knowledge engineering (KE) \citep{stranieri2001argumentation} used an adapted method of the general field argumentation model by Toulmin. The Toulmin scheme uses codes like \textit{claim, warrant, rebuttal}, and so on.
For the KE project, the Toulmin scheme was used with variation to fulfil the needs of KE practices \citep{stranieri2001argumentation}. However, another argumentation scheme dedicated to argumentation in KE was introduced by \cite{vrandecic2005diligent,pinto2004diligent}. The Diligent process model proposed an argumentation scheme in the form of an ontology using codes, such as \textit{issue, elaboration, counter example}, and so on. The argumentation ontology was focused on the ontology building process, where participants exchange arguments which may support or object to certain ontology engineering decisions \citep{tempich2005argumentation}. \cite{tempich2007argumentation} tested in practice the argumentation ontology with a group working on an ontology development, showing that it ensured agreement, helped non-experts to quickly follow processes, and made it easier to detect conflicted arguments.

In addition, prior work explored participants' roles in argumentation. Similar to the classification of arguments, several studies in Wikipedia focused on the participants' roles based on knowledge sharing like \textit{knowledge shaper, reflective reframing [reframe], reflective reinforcing [reinforcer]}, and so on \citep{hara2016co,hara2017analysis}.  
\cite{jain2014corpus} suggested a scheme of social roles in order to identify stubbornness, sensibility, influence, and ignorance in Wikipedia with roles like \textit{leader, follower, rebel}, and so on. The authors argued that different participants' roles in discussions may have an influence on the outcome of decisions and identifying the roles helps in monitoring and managing the communities.

Another view of studying argumentation is the persuasion of arguments. Under this perspective, studies have shown that convincing arguments had specific linguistic characteristics \citep{tan2016winning,wei2016post,durmus2019corpus,habernal2016makes}. Studies have investigated automated methods such as machine learning techniques, to distinguish persuasive from non-persuasive arguments. While patterns of interaction gave good results, language was essential for the classification of persuasive arguments \citep{tan2016winning}. However, in successful arguments, peoples' characteristics such as debating experience, prior success in persuasion and social network features outperformed well structure in terms of language characteristics \citep{durmus2019corpus}. In addition, argumentation characteristics such as connecting words, modal verbs, argument relevance and originality performed well in the early stages of the discussion. As the discussion progresses, responses deviated from the main subject with participants responding with less well-structured arguments \citep{wei2016post}.

\textbf{The different structure of peer-production projects}. Previous studies on disagreements in peer-production projects have suggested that the different structures of the projects, for example, an encyclopedia, a software, and a KG, may raise different challenges and indicate alternative results \citep{filippova2016effects,kittur2010beyond}. \cite{filippova2016effects} have highlighted that the different nature of peer-production projects may affect how disagreements emerged and were managed, while \cite{kittur2010beyond} found that conflict management was related to the communities' size. In addition, a study from Wikipedia Foundation by \cite{wikimediaContoversies} suggested that when comparing Wikipedia and Wikidata projects the difference in editing process needs to be considered. While many studies explored disagreements in peer-production projects, no prior work has investigated how disagreements arose, evolved, and impacted decision-making in peer-production projects that build KGs. Our study fills a gap in collaborative KGs by analysing discussions and particularly disagreements to explore collaboration and decision-making.

\section{Exploratory data analysis}
\label{sec:data}

In this work, we considered three sources of data to study disagreements: the revisions of item content pages; the item and property talk pages; and the communication pages, as shown in Table \ref{tab:list_of_discussion_channels}. In this section, we present a preliminary analysis of the revisions and the talk pages to understand how to identify disagreements in Wikidata.

\begin{figure}[h]
\centering
\includegraphics[width=\linewidth]{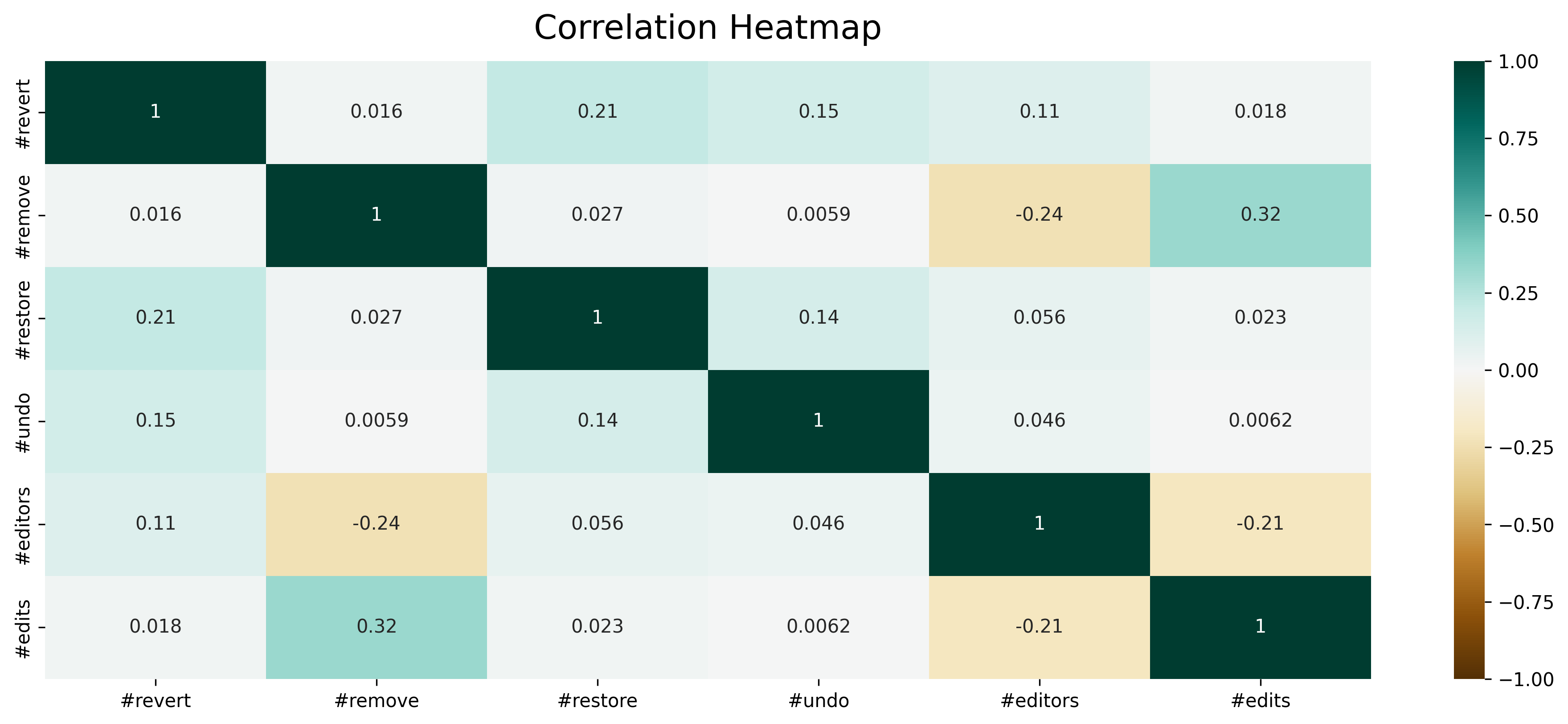}
\caption{Heatmap with the number of different types of revisions, the number of editors, and edits per item.}
\label{fig:heatmap_revisions}
\end{figure}

In Wikipedia, intense disagreements have been previously identified using revisions \citep{Borra2015, kittur2007he, rad2012identifying}, specifically reverting actions like revert or remove. We randomly sampled $0.5\%$ ($567,220$ items) of the total number of items to explore similar patterns in Wikidata. In Wikidata, we observed that the revision types \textit{revert, remove, restore,} and \textit{undo}, presented $0.05\%$, $5\%$, $0.02\%$, and $0.06\%$ respectively of the total number of revisions in the item pages we examined ($11,153,126$ revisions). 
This means that each item had a very low frequency of an average number of $0.01$ revert, $1$ remove, $0.01$ restore, and $0.02$ undo revisions. 
Moreover, the heat map of the four revision types, the number of editors, and the total number of edits per item in Figure~\ref{fig:heatmap_revisions} showed that none of these features presents a negative or positive correlation with each other. 
Hence, the type of revision was not correlated with either the amount of work or the number of people involved in the same item. Similar results were presented in a study by Wikipedia Foundation \citep{wikimediaContoversies}, where authors also noticed that reverting actions were more associated with users' characteristics than the items' content. 
In our case, from the analysis of items' revisions, it was clear that disagreements cannot easily be identified using revisions in Wikidata. 
Therefore, it was necessary to explore further the disagreements in text-based asynchronous discussions, threads, taking place on talk pages and communication pages (see Table \ref{tab:list_of_discussion_channels}).

In our previous study bout Wikidata discussions, we found that only $0.02\%$ of items had a corresponding talk page \citep{koutsiana2023ke}; this contrasts with Wikipedia where $27\%$ of the articles had one \citep{laniado2012emotions}. Similar findings have been introduced by the Wikimedia Foundation study \citep{wikimediaContoversies}. For the property talk pages, we found that despite $94\%$ of properties having a corresponding talk page, most of them included tables with information about the use of the property and only $24\%$ included discussions.
In our study \citep{koutsiana2023ke}, to understand the discussions in item and property talk pages, we randomly selected a sample of $70$ items (out of $23,271$) and $70$ properties (out of $8,406$), with various numbers of topics and discussion sizes in their talk pages. Our analysis revealed a low frequency of disagreement ($7\%$ for items and $6\%$ for properties of the codes used in the thematic analysis) and intensified conflict ($0.3\%$ for items and $0.1\%$ for properties) between the editors. 
Furthermore, a previous study showed that $95\%$ of items in Wikidata were edited by only one editor \citep{wikimediaContoversies}, indicating that there is no disagreement in editing those items. Therefore, we concluded that disagreements are very rare in the item and property talk pages and focused our analysis of disagreements on the communication pages.

To study the communication pages, we collected all Wikidata discussion pages publicly available until November $2021$. We classified the threads based on their discussion channel, such as \textit{item, property, user, and so on talk pages}, \textit{Project chat, Requests for comment}, and so on communication pages (see Table \ref{tab:list_of_discussion_channels}).
After splitting into threads, for each thread, we retained the information on the number of posts, the unique editors participating in the discussion, and the timestamps, to be able to measure the size, duration and participation levels of each discussion. Table \ref{tab:dp-attributes} presents an account of threads, posts, and unique editors for each discussion channel. 
Our complete dataset of Wikidata discussions includes more than $480K$ threads and $1M$ posts. The size of the dataset and the different discussion channels led us to use a sample of discussions for our analysis. We analysed disagreements through controversies and argumentation. Our methodology and results are presented in detail in Section~\ref{sec:methodology} and Section~\ref{sec:results}, respectively.

\begin{table}[h]
\caption{Number of threads, number of posts, and number of unique editors writing in each discussion channel from Table~\ref{tab:list_of_discussion_channels}.}
\scriptsize
\label{tab:dp-attributes}
\centering
\begin{tabular}{|p{3cm}|p{1cm}|p{1cm}|p{2cm}|}
\hline
\textbf{Discussion Channel} & \textbf{\# threads} & \textbf{\# posts} & \textbf{\# unique editors} \\ \hline
Item talk pages             & 16,510              & 33,266            & 6,101                      \\ \hline
Property talk pages         & 5,833               & 20,869            & 1,942                      \\ \hline
Users talk pages            & 84,510              & 218,994           & 11,387                     \\ \hline
Wikiproject talk pages      & 5,324               & 21,967            & 1,620                      \\ \hline
Other talk pages            & 7,549               & 27,070            & 2,493                      \\ \hline
Project chat                & 14,924              & 80,211            & 4,136                      \\ \hline
Requests for comment        & 278                 & 10,604            & 888                        \\ \hline
Report a technical problem  & 1,967               & 8,583             & 696                        \\ \hline
Request a query             & 2,414               & 9,304             & 698                        \\ \hline
Interwiki conflicts         & 3,090               & 6,602             & 2,005                      \\ \hline
Bot requests                & 1,115               & 5,705             & 552                        \\ \hline
Property proposal           & 10,009              & 76,573            & 3,058                      \\ \hline
Administrators' noticeboard & 6,522               & 27,939            & 2,258                      \\ \hline
Translators' noticeboard    & 241                 & 903               & 135                        \\ \hline
Bureacrats' noticeboard     & 301                 & 1,191             & 242                        \\ \hline
Requests for deletions      & 316,469             & 681,842           & 9,935                      \\ \hline
Properties for deletion     & 538                 & 6,775             & 739                        \\ \hline
Requests for permission     & 3,494               & 19,358            & 1,800                      \\ \hline
                  
\end{tabular}
\end{table}

\section{Methodology}
\label{sec:methodology}

In the previous section, we presented our preliminary results on analysing the Wikidata revisions and talk pages. Our findings led us to investigate disagreements in discussions taking place on communication pages. In this section, we present our methodology. Table~\ref{tab:data} shows the methods and data we used to investigate disagreements in Wikidata. Our analysis and datasets are available on GitHub (\url{https://github.com/ElisavetK/Wikidata_disagreements/tree/main}).

\begin{table}[h]
\caption{The methods and data we use to answer our research questions.}
\scriptsize
\label{tab:data}
\begin{tabular}{|M{0.11\linewidth}|M{0.15\linewidth}|l|l|m{0.25\linewidth}|}
\hline
\textbf{Research Question} & \textbf{Method}              & \textbf{\# threads} & \textbf{\# posts} & \textbf{Discussion channel used}\\ \hline
RQ1                        & Descriptive Statistics        & 481,088         & 1,257,756 & all (see Table~\ref{tab:list_of_discussion_channels})\\ \hline
RQ1 \& RQ2                  & Thematic Analysis            & 69              & 1,163 & \textit{Property proposal, Properties for deletion} and \textit{Requests for a comment}\\ \hline
RQ3                         & Measurements of Radial Trees \& Statistical Tests & 69              & 1,163 & \textit{Property proposal, Properties for deletion} and \textit{Requests for a comment}\\ \hline
RQ4 \& RQ5                  & Content Analysis            & 21              & 766 &\textit{Property proposal, Properties for deletion} and \textit{Requests for a comment}\\ \hline
\end{tabular}
\end{table}

\subsection{$RQ1$ - Where do we find disagreements in Wikidata discussions?} 
\label{sec:method-RQ1}
We used descriptive statistics for the complete set of discussions in all discussion channels ($481,088$ threads), and thematic analysis for a sample of $69$ threads. 
To identify possible discussions channels including threads with intense disagreements, we used the notion of controversies (see Section~\ref{sec:related_work}), that thread including disagreements can have \textit{a high number of participants} who continuously exchange \textit{a high number of posts} with opinions. We counted the number of posts, the number of unique editors participating, and the duration of threads in all discussion channels and identified the ones including the highest average number of participants and length per thread. 
We identified three discussion channels that could possibly include disagreements - the \textit{Property proposal, Properties for deletion} and \textit{Requests for a comment} discussion channels, that amounted to 
$10,825$ threads in total (see Section~\ref{sec:resultsRQ1}). Then, we sampled those channels to identify controversial and non-controversial discussions. 

Similar to previous studies in online discussions \citep{hara2010cross,schneider2010content,viegas2007talk}, we used 
Cochran's formula ($90\%$ confidence level and $10\%$ margin error) \citep{israel1992determining} for the sample set. The size of this sample was calculated to be $69$. As such, we sampled $23$ threads from each of the three channels. The size of our samples is quite similar to previous qualitative studies for Wikipedia that sampled approximately $25$ talk pages from article categories related to the language \citep{hara2010cross}, the page characteristics (number of contributors, number of views, and so on)\citep{schneider2010content}, and the topic (controversial or not) \citep{viegas2007talk}.

Furthermore, we classified the threads as controversial and non-controversial using qualitative techniques. 
In this study, we used two qualitative methods for text analysis: thematic and content analysis. In our analysis, one coder read the discussions and assigned themes to annotate text as summary markets. In thematic analysis, the goal was to qualify the data in order to extract a thematic map \citep{marks2004research}, while in content analysis, the goal was to quantify the data based on targeted themes \citep{vaismoradi2013content}.
We chose thematic analysis for $RQ1$ and $RQ2$ to identify all possible themes discussed in the randomly selected threads, and the content analysis for $RQ4$ and $RQ5$ to investigate predefined patterns of argumentation.

To identify whether a thread is controversial or non-controversial, the coder looked in the threads for opposing and divergent opinions, emotional language and tone, and social and cultural sensitivity.

\subsection{$RQ2$ - What are the main issues in controversial discussions in Wikidata?} 
After identifying controversial and non-controversial threads, we again used thematic analysis with the sample of $69$ threads from $RQ1$ to explore the main issues discussed. 
In collaborative KGs, the community discusses conceptual topics and KE practices, i.e., the processes, methods, languages, and tools to create, maintain, and use the knowledge \citep{studer1998knowledge}. For this reason, we started the analysis with two main themes, ``fact accuracy'' and ``KE practices''. We first identified whether the thread included a \textit{fact accuracy} or a \textit{KE} controversy. To distinguish the two themes, the coder checked: for technical KE language, such as ontology, taxonomy, constrains, and semantics (i.e., KE theme), or domain specific terms, such as biology, sports, medicine, and books (i.e., fact accuracy theme); and whether the discussion was about how to organise and represent knowledge (i.e., KE theme), or facts, practices, and decisions within a field (i.e., fact accuracy theme). Next, we determined the specific cases, such as disagreement among editors on what should be done with references that did not have valid websites. Finally, we classified the specific cases into themes, such as sources, policies, quality, and so on, to present the main issues of controversy in Wikidata discussions.

\subsection{$RQ3$ - What are the characteristics of controversial discussions in Wikidata?}
After investigating the main issues in controversial threads, we continued by studying their characteristics.
We used the sample of $69$ threads from $RQ1$ to study whether there was a correlation between various features of controversial and non-controversial threads. We used Spearman's rank coefficient \citep{zar2005spearman} for nonparametric variables to calculate correlation for three types of features: thread measurements, i.e., the number of posts, the number of editors, and the duration of the discussion; the label of controversial and non-controversial identified in $RQ1$; and measurements of \textbf{radial trees}.

\begin{figure}[t]
\centering
\includegraphics[width=0.5\linewidth]{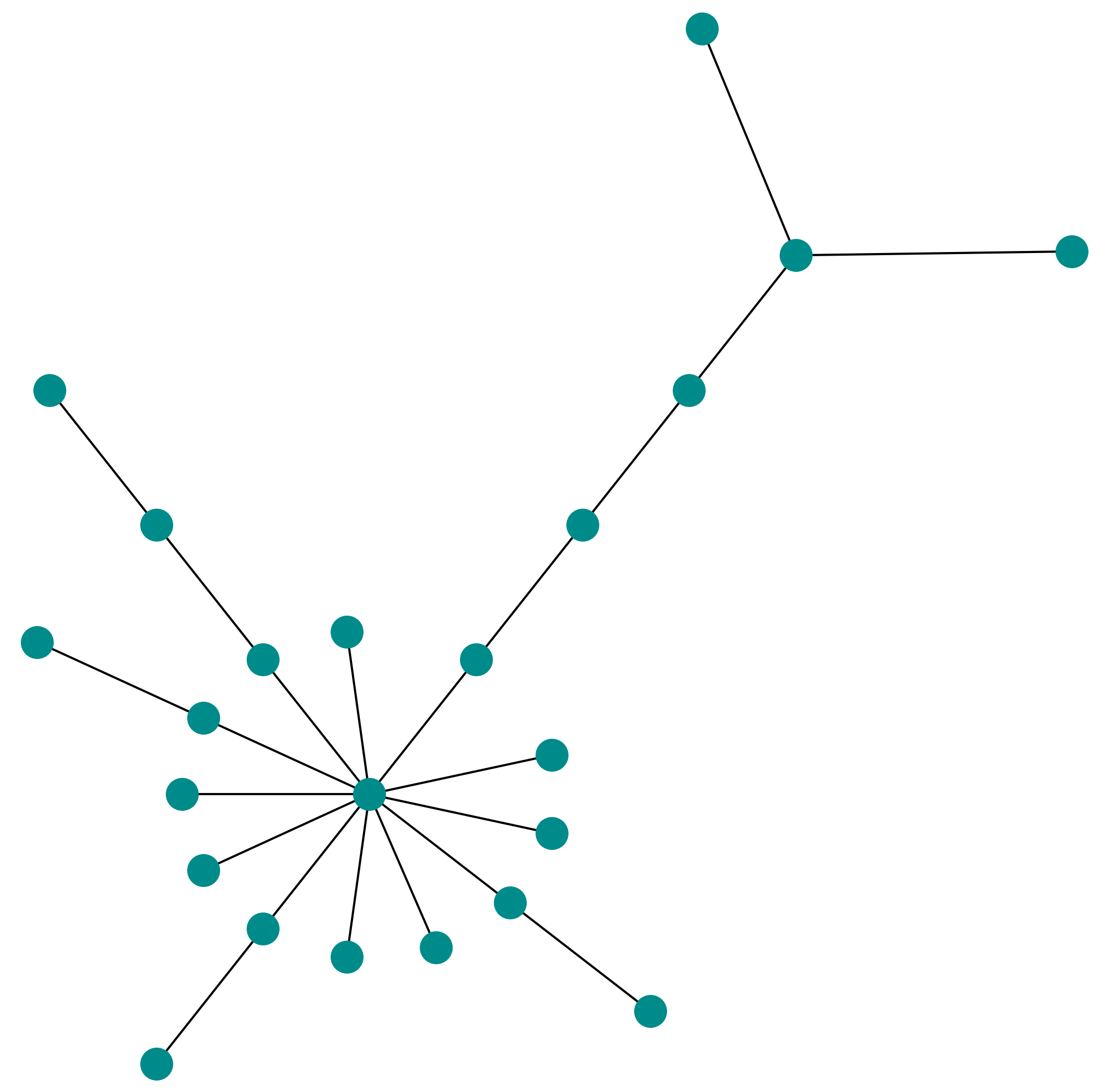}
\caption{A radial tree for a thread discussion.}
\label{fig:radial-tree}
\end{figure}

A radial tree \citep{herke2009radial} represents each thread in a circular tree structure. We considered the first post, stating the main issue of the thread, as the central node. Any post replying directly to the first post was considered as the first nested level. Respectively, any post replying to other posts expanded the tree by increasing the nested levels. After including all the posts in a thread, we obtained a circular structure that represents the direction of editors' replies in a discussion radially (Figure~\ref{fig:radial-tree} shows examples of a radial tree).
Using this tree representation, we were able to study the hierarchy of discussions and the nested patterns. 

Similar to previous studies on Slashdot \citep{gomez2008statistical} and Wikipedia \citep{laniado2011wikipedians} discussions, we used two measurements of radial trees for our analysis: the h-index \citep{hirsch2005index}, and the depth of the tree.
We measured an adapted version of h-index (commonly used to measure researchers' scientific output) proposed by \cite{gomez2008statistical}. The h-index is defined as the maximum rank number when the number of citations was greater than or equal to h. 
In this study, we used an adapted metric, where given a radial tree for a thread, we counted the number of radii per nested level, starting with root node i.e. level one. 
The h-index was then determined to be the maximum level for which the number of radii was greater than or equal to the level number.
For example, for the radial tree in Figure~\ref{fig:radial-tree}, the first level is the root node, the second level has $12$ radii, the third $5$, the fourth $2$, and so on and so forth, giving an h-index of $2$. 
Furthermore, we measured the depth, i.e., the maximum number of levels for the radial trees to investigate the sub-threads in the discussions. This turned out to be $6$ for the previous example in Figure~\ref{fig:radial-tree}.

\subsection{$RQ4$ - How do editors argue when they disagree in Wikidata?}
After the study of controversial threads we continued our analysis by exploring argumentation.
We investigated the argumentation techniques used in controversial discussions to express opinions.
We investigated the argumentation techniques in controversial discussions identified in $R1$ using content analysis \citep{vaismoradi2013content}.

Table \ref{tab:ca} presents the coding scheme we used to identify patterns of augmentations and answer $RQ4$. Previous coding schemes in knowledge collaboration projects like Wikipedia \citep{hara2016co,freard2010role} have focused on knowledge sharing. Furthermore, similar studies in open source software development \citep{wang2020argulens} and KE \citep{stranieri2001argumentation} projects used a general field argumentation model by Toulmin with several variations to adjust for their needs, such as ontology development. However, we chose to use the Diligent argumentation framework \citep{tempich2007argumentation}, which is focused on the argumentation patterns in collaborative KE. The Diligent framework has been specifically designed for collaborative KE projects with main concepts being \textit{issues, ideas} and \textit{arguments}. The \textit{issue} introduces the initial subject for discussion. An \textit{idea} is a response to an issue. Moreover, \textit{arguments}, positive or negative, could \textit{justify} or \textit{challenge} an issue or ideas. Furthermore, positions could \textit{agree} or \textit{disagree} with issues, ideas, or arguments. Finally, the discussion state is shown by a decision, which could be still \textit{open} and \textit{under discussion}, or \textit{postponed, discarded} or \textit{agreed}. With this coding scheme, we identified types of arguments in controversial threads and we continued our analysis with the role of editors in argumentation. 

\begin{table}[h]
\caption{Coding scheme used for the content analysis to study the arguments based on the Diligent argumentation framework \cite{tempich2007argumentation}.}
\scriptsize
\label{tab:ca}
\begin{tabular}{|m{3cm}|m{5cm}|}
\hline
\textbf{Code/Argument}      & \textbf{Description}                                                                                                                                   \\ \hline
Issue                        & It is the main subject starting the discussion. \\ \hline
Idea                        & It is a new subject for discussion during the argumentation process. \\ \hline
Elaboration                 & It is the extension of an issue presenting additional details.                                                                                         \\ \hline
Justification - Example     & Justification is an argument supporting an idea or issue. Example is used to provide evidence to increase the belief on a matter.                      \\ \hline
Justification - Evaluation  & Justification is an argument supporting an idea or issue. Evaluation gives a measurable mean to increase belief on a matter.                           \\ \hline
Challenge - Counter Example & Challenge is an argument against an idea or issue. Counter Example provides counter evidence for a matter.                                             \\ \hline
Challenge - Alternative     & Challenge is argument against an idea or issue. Alternative is used to suggest an alternative solution on a matter.                                     \\ \hline
Position - Agree            & This clarifies the participant's position on an issue, idea,  or argument, either by voting support or by agreeing within the sequence of arguments.    \\ \hline
Position - Disagree         & This clarifies the participant's position on an issue, idea,  or argument, either by voting oppose or by disagreeing within the sequence of arguments. \\ \hline
Decision - Under discussion & The issue is still open and anyone can argue.                                                                                                          \\ \hline
Decision - Postponed        & There was no consensus on the issue, but it is frozen and no one can express a new opinion on it.                                                        \\ \hline
Decision - Discarded        & The initial issue or an idea was rejected as not relevant.                                                                                             \\ \hline
Decision - Agreed           & The final decision has been taken regarding the initial issue or an idea that was formalised over the sequence of arguments.                           \\ \hline
\end{tabular}
\end{table}



\subsection{$RQ5$ - What are the roles of editors in argumentation in Wikidata?} 
Similar to $RQ4$, we used content analysis for the threads identified as controversial in $RQ1$ to identify the role of editors in argumentation.
The coding scheme used to study the roles of the different participants is presented in Table \ref{tab:actors}. This scheme was based on a study by \cite{jain2014corpus} that identified social roles in contentious online discussions. 
Other similar studies about the content of arguments were focused on the roles based on knowledge sharing and modifying, and facilitating collaboration, such as knowledge shaper, organiser, giver, and so on \citep{faraj2011knowledge,kane2014emergent,hara2017analysis}. However, our aim was to understand the roles of social interactions. We used the roles suggested by \cite{jain2014corpus} with small variations to the names as \textit{leader, follower, rebel, ignored rebel, outsider, loner, disruptor}. This coding scheme helped us to investigate types of editors in argumentation in the sense of whether we could find editors that influence others, editors whose opinions are ignored, editors who have minimum participation, or spammers and trolls. 

\begin{table}[h]
\caption{Coding scheme used for the content analysis to study the roles of participants in argumentation.}
\label{tab:actors}
\scriptsize
\begin{tabular}{|m{2cm}|m{6cm}|}
\hline
\textbf{Participant role}      & \textbf{Description}                                                                                                                                                                                                      \\ \hline
Leader              & An editor who manages to influence another editor to change her/his opinion or influence her/him to support her/his arguments.                                                                                              \\ \hline
Follower            & An editor who is influenced by another editor and changes her/his opinion or supports the other editor's argument usually follows the ``Leader''. She/he can be an editor who does not have her/his own arguments. \\ \hline
Rebel               & An editor who drives the discussion in some direction. She/he is devoted to the discussion and engaged with other editors. A ``Rebel'' posts sensible arguments which cannot be ignored by others.                    \\ \hline
Ignored Rebel & An editor with similar behaviour as a ``Rebel''. The main difference is that hers//his arguments are ignored by others.                                                                                                  \\ \hline
Outsider           & An editor who does not contribute to the main purpose of the discussion. She/he is engaged in the discussion but the content is either emotional, illogical, or out of topic.                                                 \\ \hline
Loner    & An editor who makes minimal but legitimate contributions to the discussion.                                                                                                                                                \\ \hline
Disruptor           & An editor who makes minimal and unuseful contributions to the discussion. She/he can be a potential spammer.                                                                                                              \\ \hline
Others           & For editor who do not follow any of the previous behavioural patterns.                                                                                                             \\ \hline
\end{tabular}
\end{table}


\section{Results}
\label{sec:results}

In this section, we present the insights from our analysis on discussions towards the different research questions.

\subsection{$RQ1$: Where do we find disagreements in Wikidata discussions?}
\label{sec:resultsRQ1}

\paragraph{Summary} We suggest that \textit{Properties for deletion, Property proposal}, and \textit{Requests for comments} discussion channels are the ones that are most likely to include threads with disagreements. For a sample of threads randomly selected from the three discussion channels, we identified $30\%$ as controversial.

\begin{figure*}[h]
    \centering
    \subfigure[]{\includegraphics[width=0.48\textwidth]{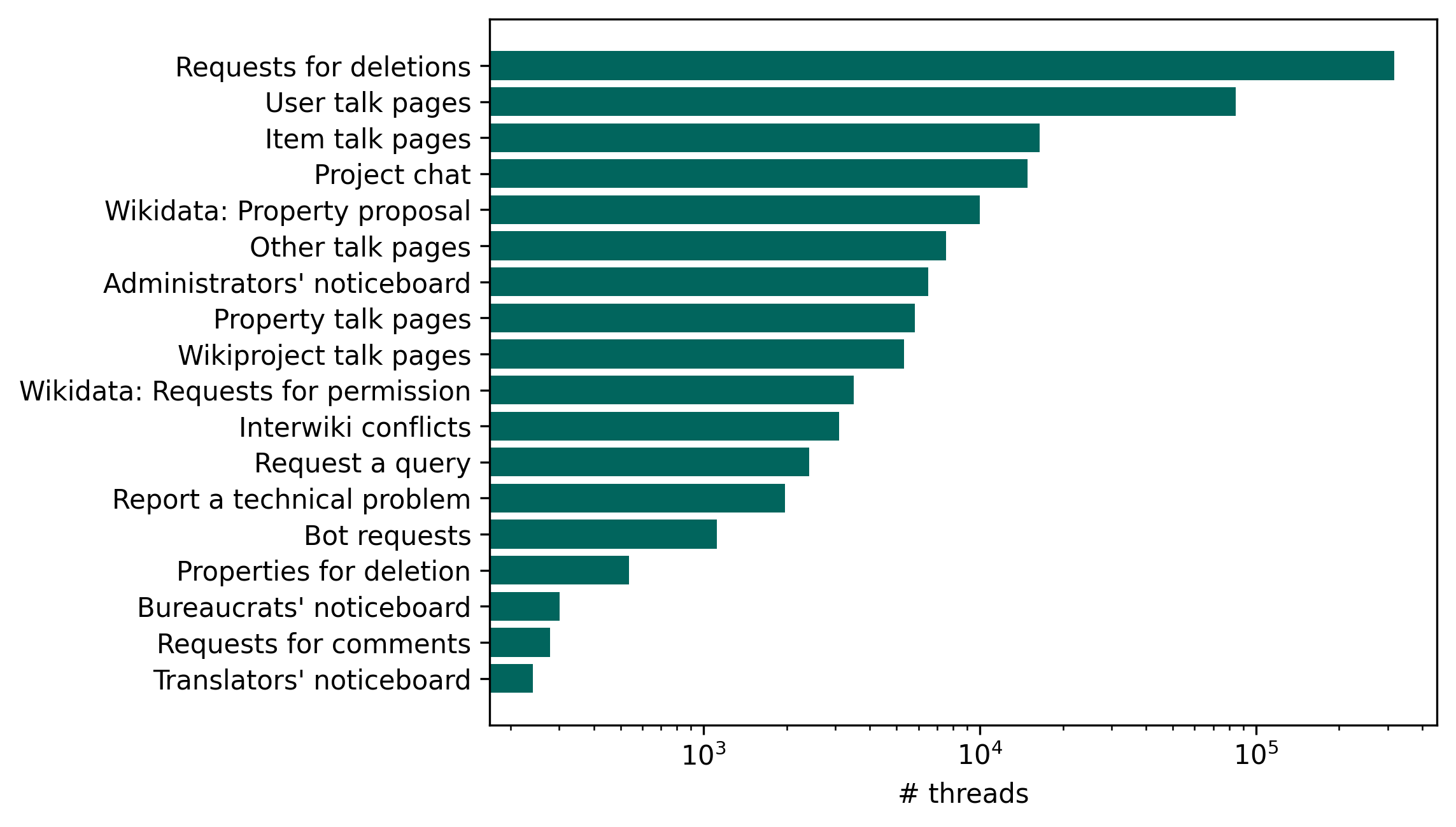}} 
    \subfigure[]{\includegraphics[width=0.48\textwidth]{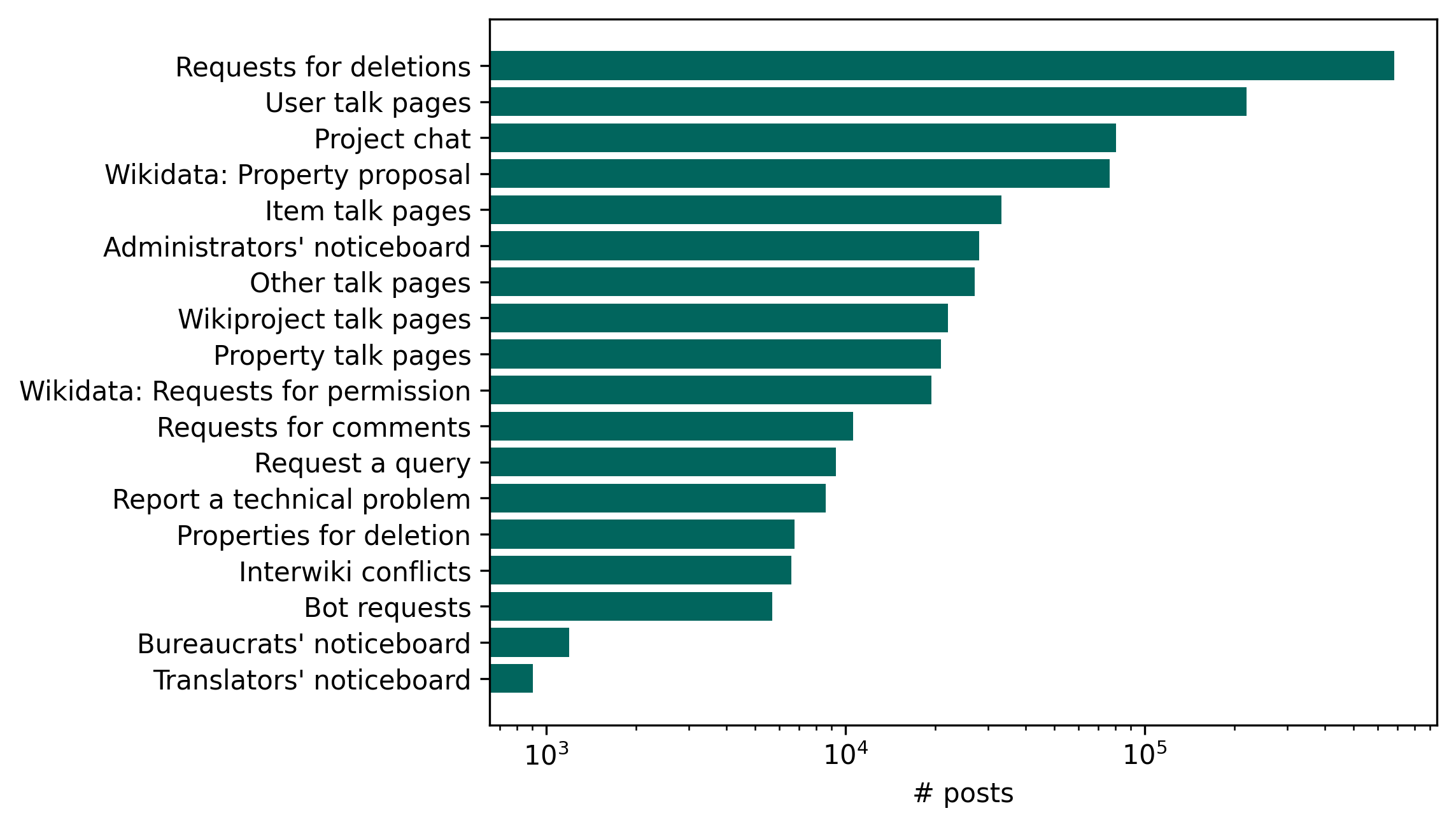}} 
    \caption{(a) the number of threads and (b) the number of posts for the different discussion channels.}
    \label{fig:stats1}
\end{figure*}

Our analysis shows that the discussion channel with the highest traffic (i.e., the most visited channel with the highest number of threads) is \textit{Requests for deletion}, presenting $66\%$ of the total threads ($316,469$ threads, and $681,842$ posts). This is followed by the \textit{User talk page} channel with $18\%$ of the total threads ($84,510$ threads, and $218,994$ posts). Figure~\ref{fig:stats1} shows (a) the number of threads and  (b) the number of posts for the different discussion channels.

\begin{figure*}[h]
    \centering
    \subfigure[]{\includegraphics[width=0.48\textwidth]{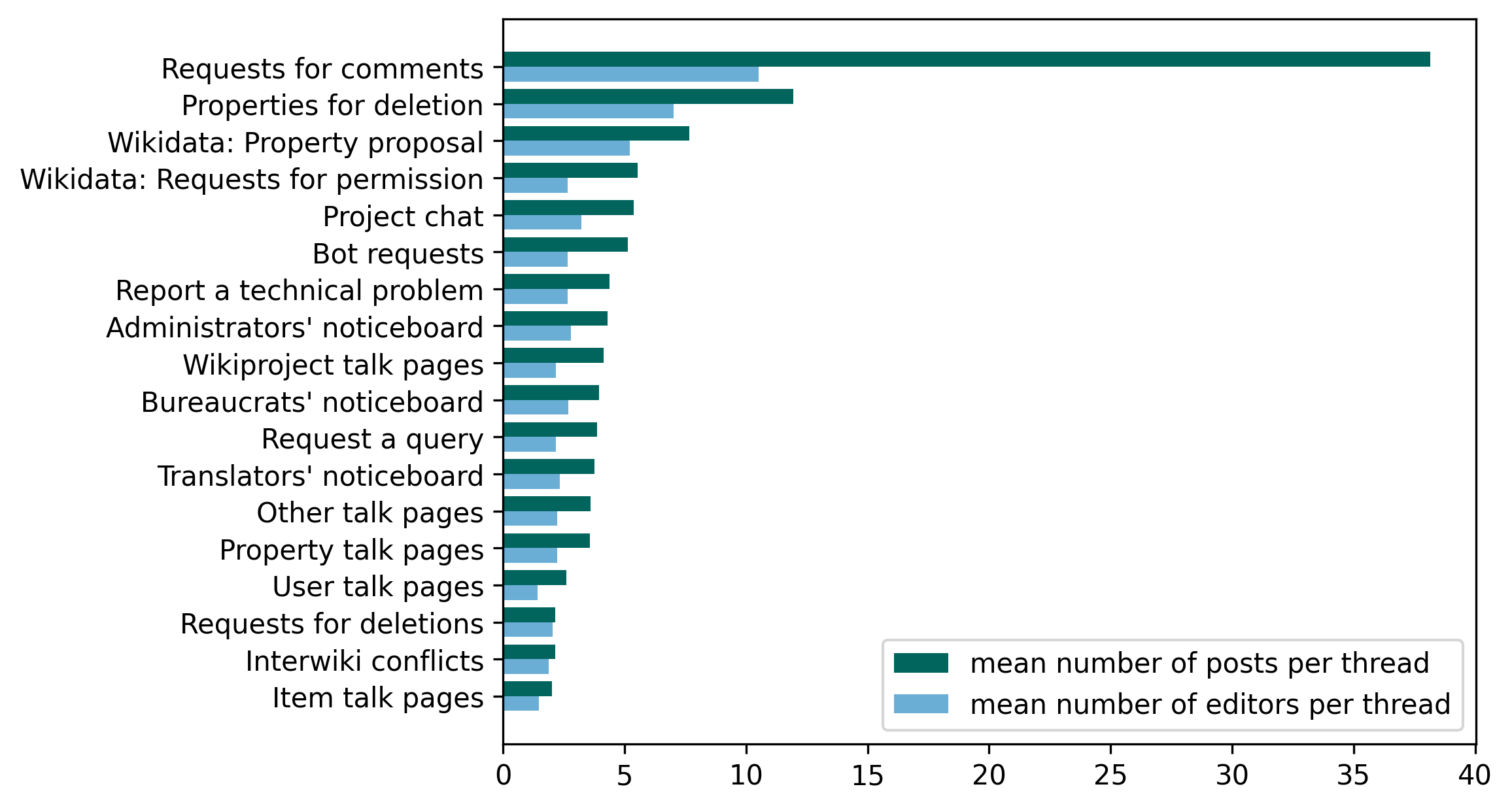}} 
    \subfigure[]{\includegraphics[width=0.48\textwidth]{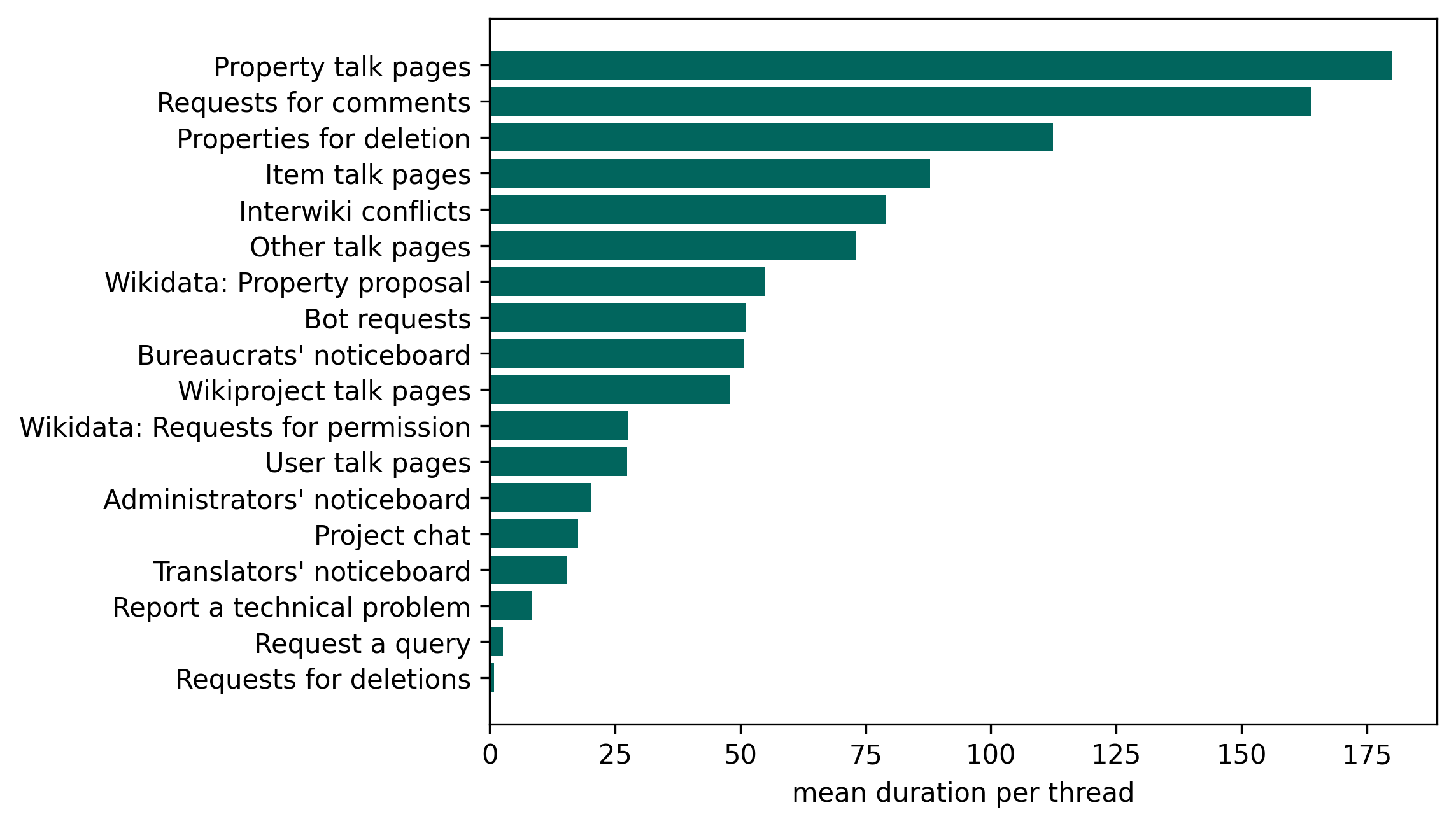}}
    \caption{(a) the mean number of posts and editors per thread and (b) the mean duration of threads  for the different discussion channels.}
    \label{fig:stats3}
\end{figure*}

Based on our notion of intense disagreements (see Section~\ref{sec:related_work} and Section~\ref{sec:methodology}) potential threads with disagreements can include \textit{a high number of posts} and \textit{a high number of participants}. While \textit{Requests for deletion} and \textit{User talk page} are the most visited channels, they host very short threads with an average number of posts per thread of $2$ and $3$ respectively (see Figure~\ref{fig:stats3} (a)) and very short duration (Figure~\ref{fig:stats3} (b) shows one day for \textit{Requests for deletion}). These short, quick discussions made for the conversation difficult to evolve into a disagreement. Figure~\ref{fig:stats3} (a) shows that \textit{Properties for deletion, Property proposal}, and \textit{Requests for comments} present the highest number of posts and editors per thread in Wikidata indicating that they may include controversial discussions.

As mentioned in Section~\ref{sec:methodology}, to investigate this hypothesis, we first sampled the threads from the three channels based on Cochran's Formula, which resulted in $69$ threads in total ($23$ per discussion channel).
With thematic analysis, $30\%$ ($21$ out of $69$) of threads were assigned to be controversial. 
To be specific, we identified $30\%$ of the threads at \textit{Properties for deletion} ($7$ out of $23$), $9\%$ of threads at \textit{Property proposal} ($2$ out of $23$), and $52\%$ of the threads  at \textit{Requests for comments} ($12$ out of $23$) as controversial. We found that \textit{Requests for comments}, with a mean number of $38$ posts and $11$ editors per thread, is the most controversial among the three discussion channels that were analysed. However, for \textit{Property proposal}, with a mean number of $8$ posts and $5$ editors per thread, we found only $2$ controversial threads.

The results suggest that a high number of posts combined with a high number of editors in threads could indicate controversial discussions. However, we conjecture that despite the length of the thread, the content of the discussion and the policy of use for each discussion channel have an essential role in controversies and their evolution. For this reason, we further analysed the controversial threads to identify the main issues of controversy in Wikidata.  

\subsection{$RQ2$: What are the main issues in controversial discussions in Wikidata?}
\label{sec:RQ2}

\paragraph{Summary} We found that the main issues in controversial discussions are related to KE ($95\%$ of the threads) and particularly concern process controversies (i.e., policies or practices). Furthermore, controversies do not often lead to conflicts. We found that more than half of controversial discussions do not reach a consensus, while in the \textit{Property proposal} channel, a couple of votes are enough to take decisions on the creation of a property, contrasting the decision on the deletion of a property.

As described in Section~\ref{sec:methodology}, in collaborative KE projects, we expected to find two \textbf{main themes} for disagreements, one related to fact accuracy, and another associated with KE practices. We found that $20\%$ ($4$ out of $21$) of the controversial threads are related to fact accuracy, with $3$, including both fact accuracy and KE practices issues. The majority of threads, $95\%$ ($20$ out of $21$) of threads, are related to KE practices issues.

\begin{figure*}[h]
\centering
\includegraphics[width=0.9\linewidth]{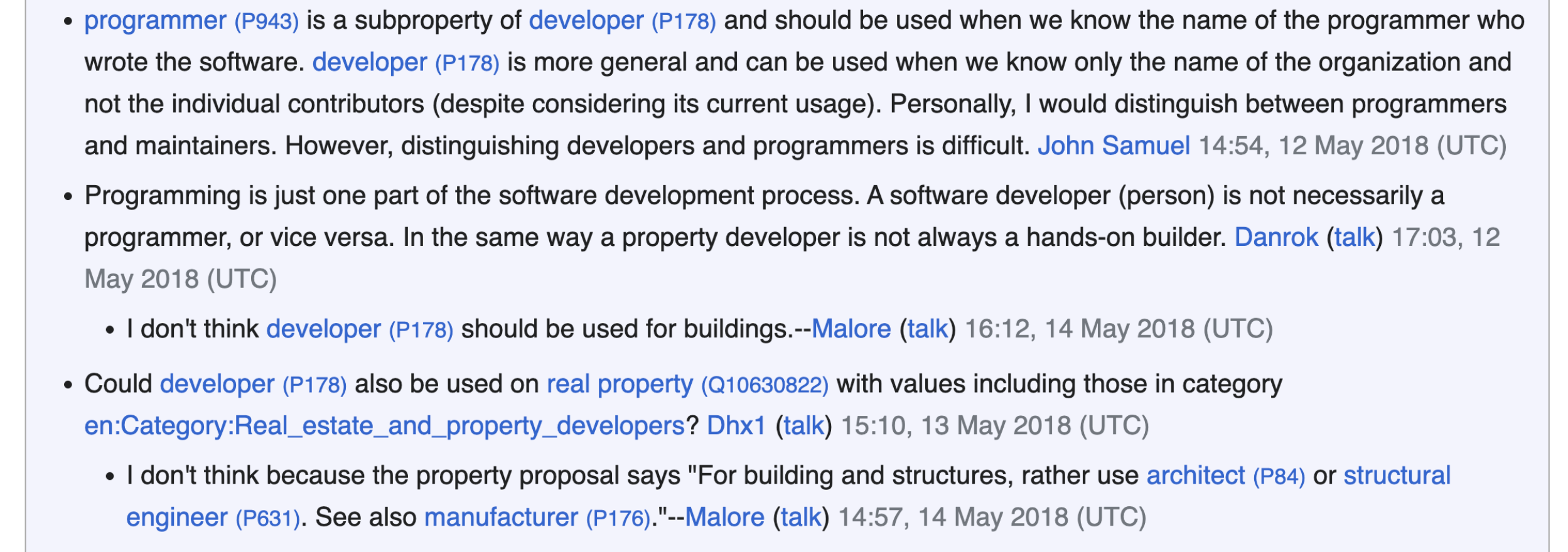}
\caption{An example discussion from Wikidata \textit{Requests for comment} discussion channel. The issue raised concerns the use of the properties ``programmer'' and ``developer'' and is an example of the theme fact accuracy.}
\label{fig:programmer}
\end{figure*}

For \textbf{fact accuracy}, we found three cases regarding the use of specific properties based on the meaning of their name. For example, this happens for the properties ``given name'', ``surname'', and ``birth name'', where editors disagreed on the appropriate property to be used in different cases. 
The properties ``developer'' and ``programmer'' also have a similar case of disagreement. Figure~\ref{fig:programmer} presents part of the discussion related to the issue about the ``developer''.
Further, a controversy evolved about the creation of a new property named ``at'' since its generic naming could cause semantic misunderstandings. 
Finally, we found a discussion about the evaluation of data quality in Wikidata, where editors disagreed on the definitions of data and information quality.

\begin{figure*}[h!]
\centering
\includegraphics[width=0.9\linewidth]{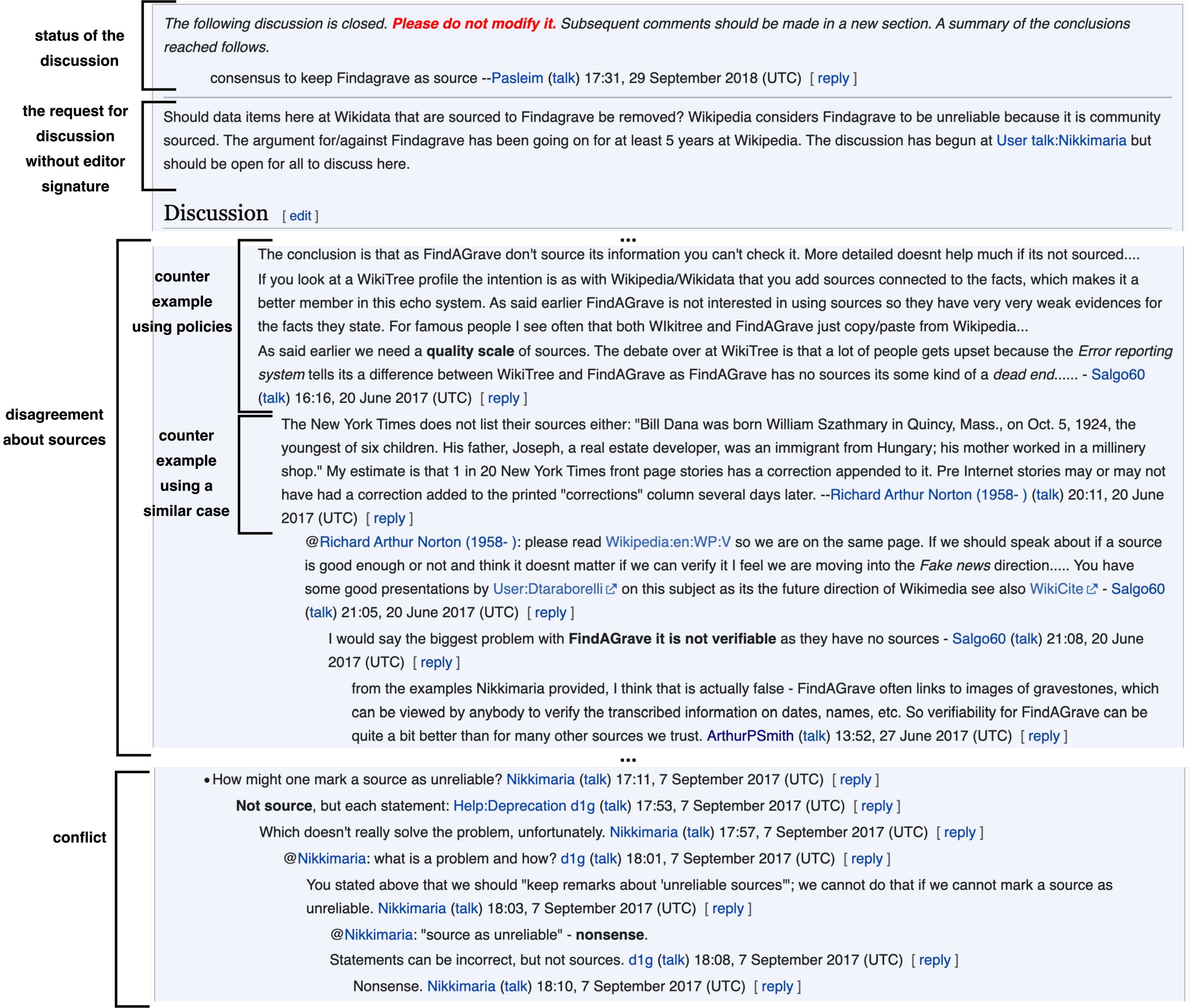}
\caption{An example discussion from Wikidata \textit{Request for comments} discussion channel. The first component of the example includes the initial request and the current status of the issue. The second component of the example shows part of the discussion with an intense disagreement and two counter example arguments. The last component of the example shows part of the discussions with a conflicting interaction.}
\label{fig:find_a_grave}
\end{figure*}

\begin{figure*}[h!]
\centering
\includegraphics[width=\linewidth]{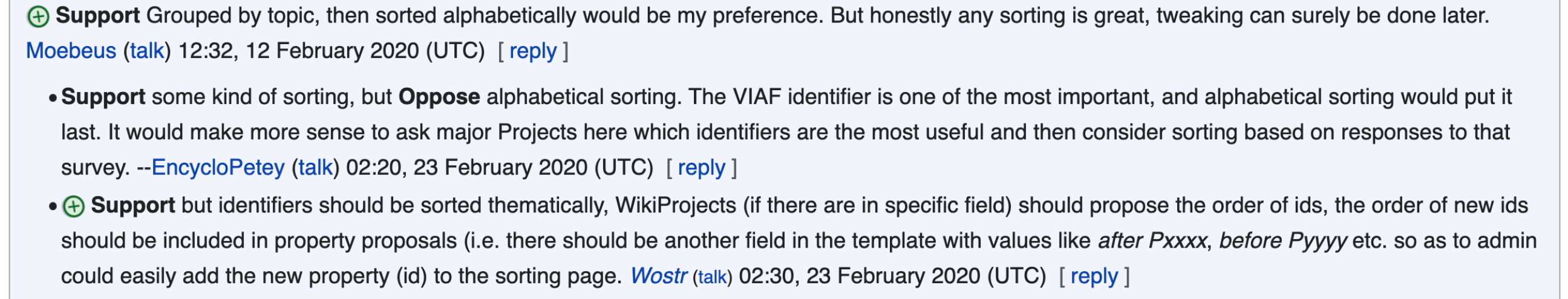}
\caption{Part of a discussion about how Wikidata should order the list of properties.}
\label{fig:properties_order}
\end{figure*}

\begin{figure*}[h!]
\centering
\subfigure[]{\includegraphics[width=\textwidth]{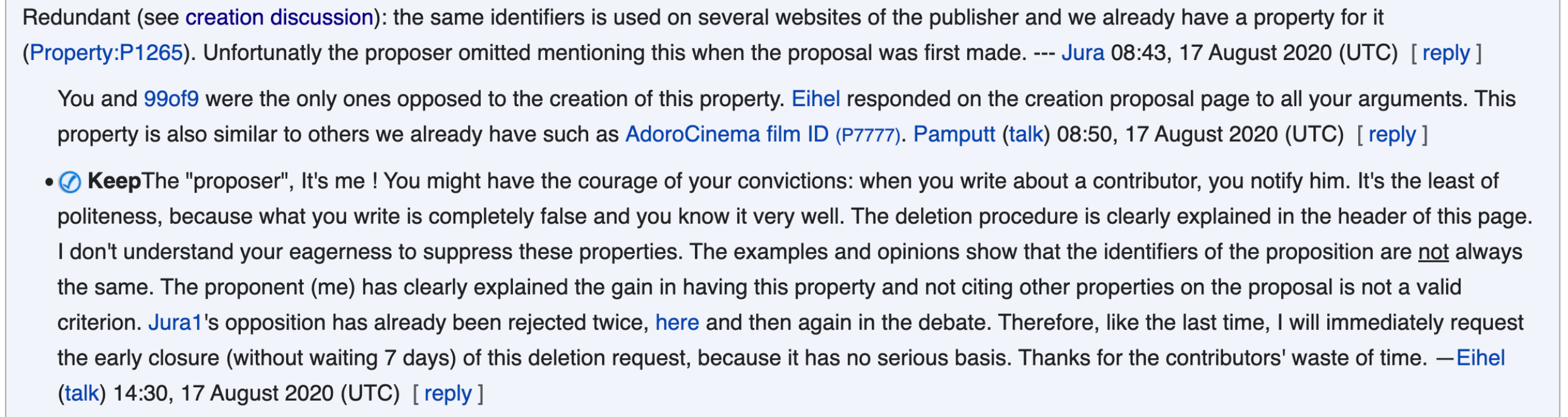}} 
\subfigure[]{\includegraphics[width=\textwidth]{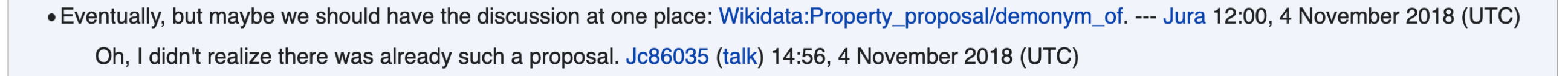}}
\caption{(a) Part of a discussion for a suggestion to delete a property. The participants often refer to discussions on other pages, especially in the property proposal of this property. (b) Part of a discussion between two editors, who are suggesting that this discussion should be held in one place.}
\label{fig:links_to_other_disc}
\end{figure*}

\begin{figure*}[h!]
\centering
\subfigure[]{\includegraphics[width=\textwidth]{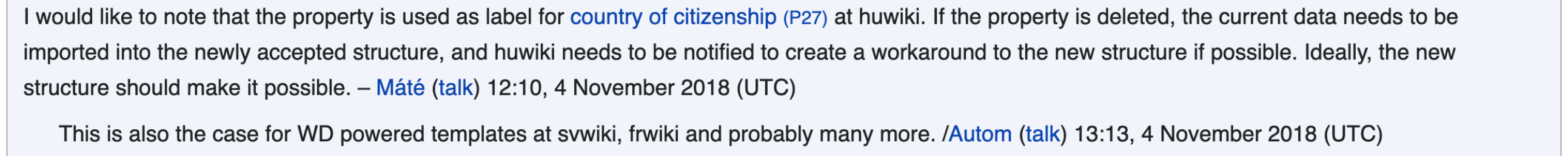}} 
\subfigure[]{\includegraphics[width=\textwidth]{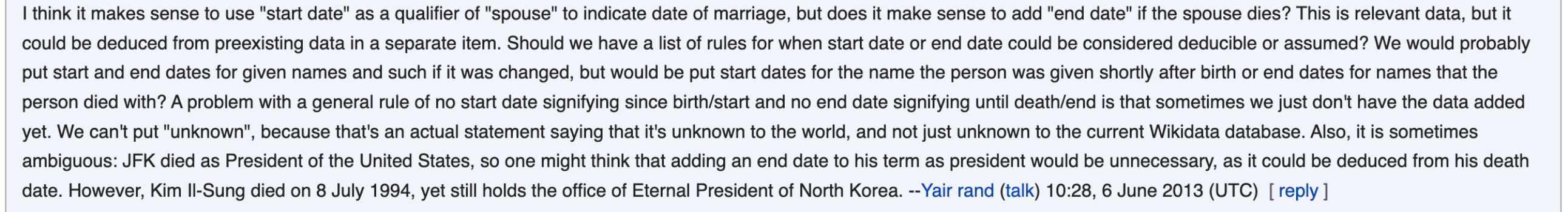}}
\caption{Examples of (a) \textit{task controversy}  related to property taxonomy, and (b) \textit{process controversy} related to qualifier guidelines}
\label{fig:controversies}
\end{figure*}

For \textbf{KE practices}, we found that \textit{Requests for comments} includes issues related to: interwiki links (i.e., links that connect the different Wikimedia projects like Wikipedia, Wikibooks, Wikisource, and so on); the references linked to the pages as justification of facts; the quality of the project; the policies and practices of Wikidata; and the use of properties. However, for the other two discussion channels, as their name and purposes indicate, we found issues related to the use of properties.
The \textit{references} linked to sources to verify the accuracy of facts connected to items. Editors disagreed on the type of sources they used to justify the facts, and the changes made over time to the source websites. Figure~\ref{fig:find_a_grave} shows an example of a controversy about references. The reference issue is the use of the source ``Find a Grave'' for the date of death. The discussion was active for $6$ months, including $73$ posts and $17$ unique editors. The editors disagreed on the reliability of this source, since it is a community-sourced project, and on its use as a reference.
Furthermore, changes in \textit{guidelines, policies} and \textit{practices} attract a variety of opinions in any peer-collaboration. In Wikidata, we found controversies about changes in guidelines regarding the creation of a property or the writing instructions in the discussion channel \textit{Request for comment}. Figure~\ref{fig:properties_order} presents another case where editors disagreed on how Wikidata should present the list of properties, for example, in alphanumeric order, in groups based on the theme, or by listing the most important properties at the top. The figure shows that editors vote \textit{support} on this issue, but they disagree on the way it should be done.
In addition, controversies about the \textit{use of properties} are related to constraints (i.e., rules for the use of the properties), qualifiers (i.e., properties used in combination with other properties to extend their semantic connections), and taxonomic use (i.e., properties connect items to classes and subclasses).
We found that often in threads about property issues, editors point to discussions about the same issues that occur in other discussion channels. Figure~\ref{fig:links_to_other_disc} (a) shows an example of a discussion related to the deletion of a property, and editors refer to the discussion related to the proposal of this property. Furthermore, Figure~\ref{fig:links_to_other_disc} (b) points out the confusion of two participants regarding discussions about the same topic in multiple places. This may suggest that discussions about properties, proposal or deletion, need attention.

Based on previous studies in peer-production about conflict categorisation \citep{filippova2015mudslinging,arazy2013stay} (see Section~\ref{sec:methodology}), we identified two \textbf{types of controversies} in the analysed threads, namely \textit{task} and \textit{process}.
In Wikidata one can think of task controversy as disagreements about the editing of an item or properties and process controversy as disagreements on how to perform a task rather than the task itself. The latter is related to policies or practices set by the community in order to preserve knowledge consistency, such as the use of the \textit{instance of (P31)} property when one cannot specify the \textit{subclass of (P279)} property in an item.

Figure~\ref{fig:controversies} (a) shows an example of task controversy and Figure~\ref{fig:controversies} (b) for process controversy from our analysis. For \textit{task controversy} ($38\%$ of threads), the issues were related mostly to fact accuracy and particularly to properties. Furthermore, for KE practices, task controversy could be seen in the taxonomy of properties, such as a disagreement for the deletion of a property due to the discontinuity between Finland and Swedish wikis. For \textit{process controversy} ($52\%$ of threads), the issues were related to KE practices. We found issues about guidelines, policies and practices, such as guidelines in links connecting the different Wikimedia projects, and policies on use of property constraints and qualifiers. The rest of the threads ($10\%$) were not recognised as task or process controversies.

\begin{figure}[h]
\centering
\includegraphics[width=0.8\linewidth]{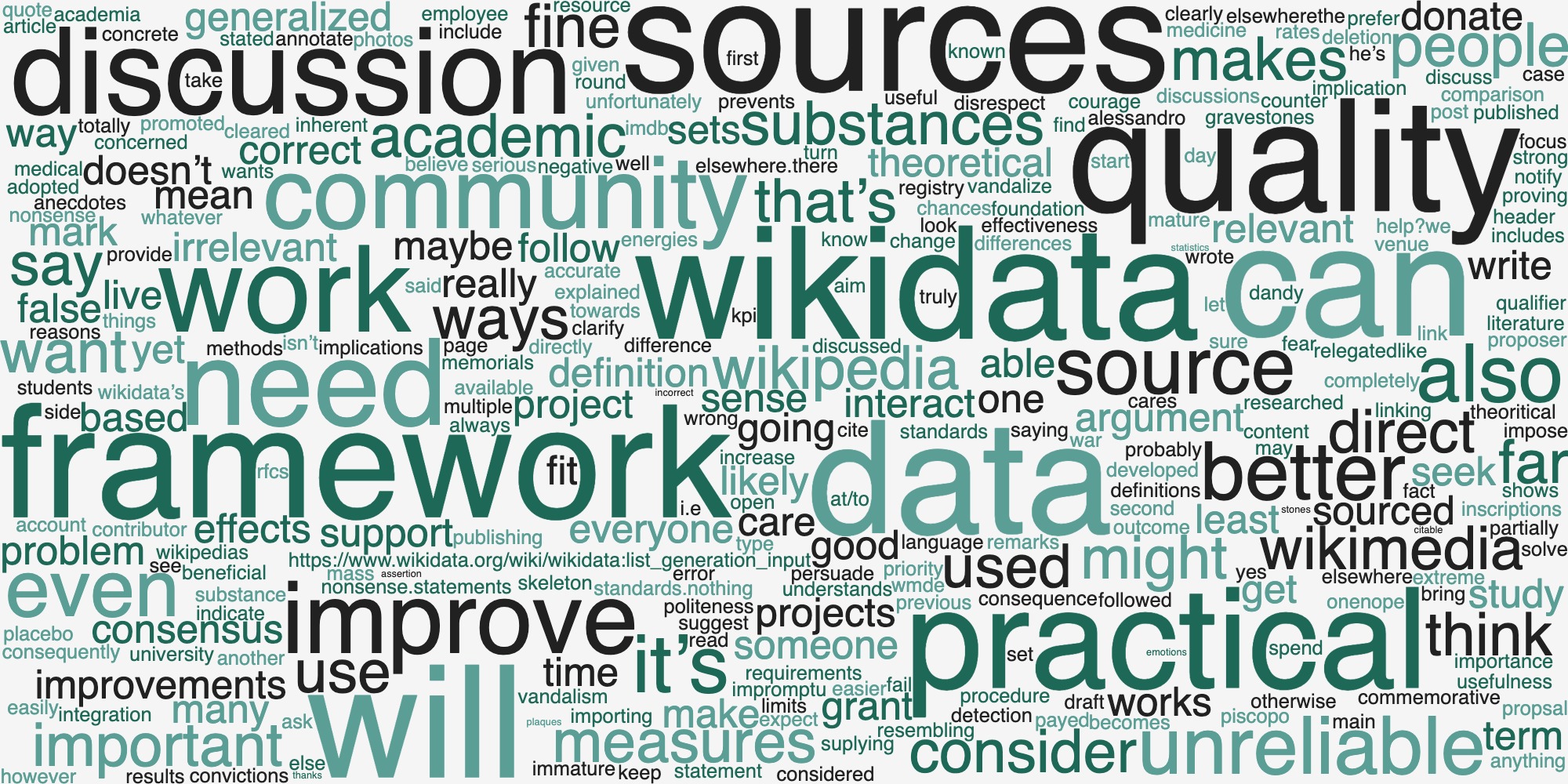}
\caption{Wordcloud for the posts identified as conflicts.}
\label{fig:conflict}
\end{figure}

While we identified controversial threads in Wikidata, we found that controversies do not lead to intense \textbf{conflicts} with use of strong language or threats. The coder distinguished controversies with conflicts when: there was not a structured disagreement, often intellectual or ideological, rather than an emotionally charged clash, often personal; there was not debate, opposing arguments, or critique, rather than tension, anger, disruption, or hostility; and there was not issue focus, rather than person or group focus. The identified conflicting posts were either ironic, indicated anger, or used expressions like ``nonsense'' and ``this is irrelevant'' (see Figure~\ref{fig:find_a_grave}). Figure~\ref{fig:conflict} presents a wordcloud for the posts we identified as a conflict. We can see that the main topics of conflict are ``sources'' and ``quality''.
$30\%$ of the threads identified as controversial included minor conflicts which lasted from $2$ to $4$ posts and arose between two editors. The other editors participating in the thread continued with their arguments and did not get involved in the conflict. In most cases, controversies in Wikidata evolve with peaceful arguments without any intense conflict. The reason behind this could be the fact that editors do not revert or change the work of others repeatedly. This is different from co-editing activities in Wikipedia, such as writing an article, which could lead to edit wars since in Wikidata each suggestion needs to reach a consensus to proceed to changes. 

Studying the decisions in our analysed threads we found that often, there is no \textbf{consensus} in controversial discussions, $62\%$ ($13$ out of $21$) of threads did not lead to consensus.
On the contrary, in non-controversial threads, $38\%$ ($18$ out of $48$) did not lead to consensus.
We found many cases where the discussion was open for months without being active, and administrators closed them without consensus.

Regarding the \textbf{non-controversial} threads, we identified two types: short threads and the threads that include many vote posts with no further justification. 
The \textit{Property proposal} was the least controversial between the three channels. 
This could be because the editors reach a consensus with $2$ to $3$ votes and thus, disagreement is more unlikely. This contrasts with the other two discussion channels, where we found extended voting and elaboration around the discussed issue. This verifies our suggestion in Section~\ref{sec:resultsRQ1} that the size of a thread is not enough, and the content and discussion channel type have an essential role in identifying controversial discussions.

\subsection{$RQ3$: What are the characteristics of controversial discussions in Wikidata?}
\label{sec:RQ3}

\paragraph{Summary} To find the characteristics of controversial threads, we used the sample of $69$ threads and the labels of controversial and non-controversial threads from $RQ1$. Based on descriptive statistics and the radial trees built for each thread, we measured the number of posts, the number of unique editors participating in the discussions, the duration of the discussion, the h-index, and the depth of the thread (i.e., the number of levels in a thread). 
We found that controversies are correlated with the number of participants ($0.7$) and posts ($0.72$), and the depth of the thread ($0.69$), but not with the duration ($0.52$) of the thread. Furthermore, in controversial discussions, it is likely to find many participants exchanging fewer posts than a few participants exchanging many posts.

\begin{figure}[h]
\centering
\includegraphics[width=\linewidth]{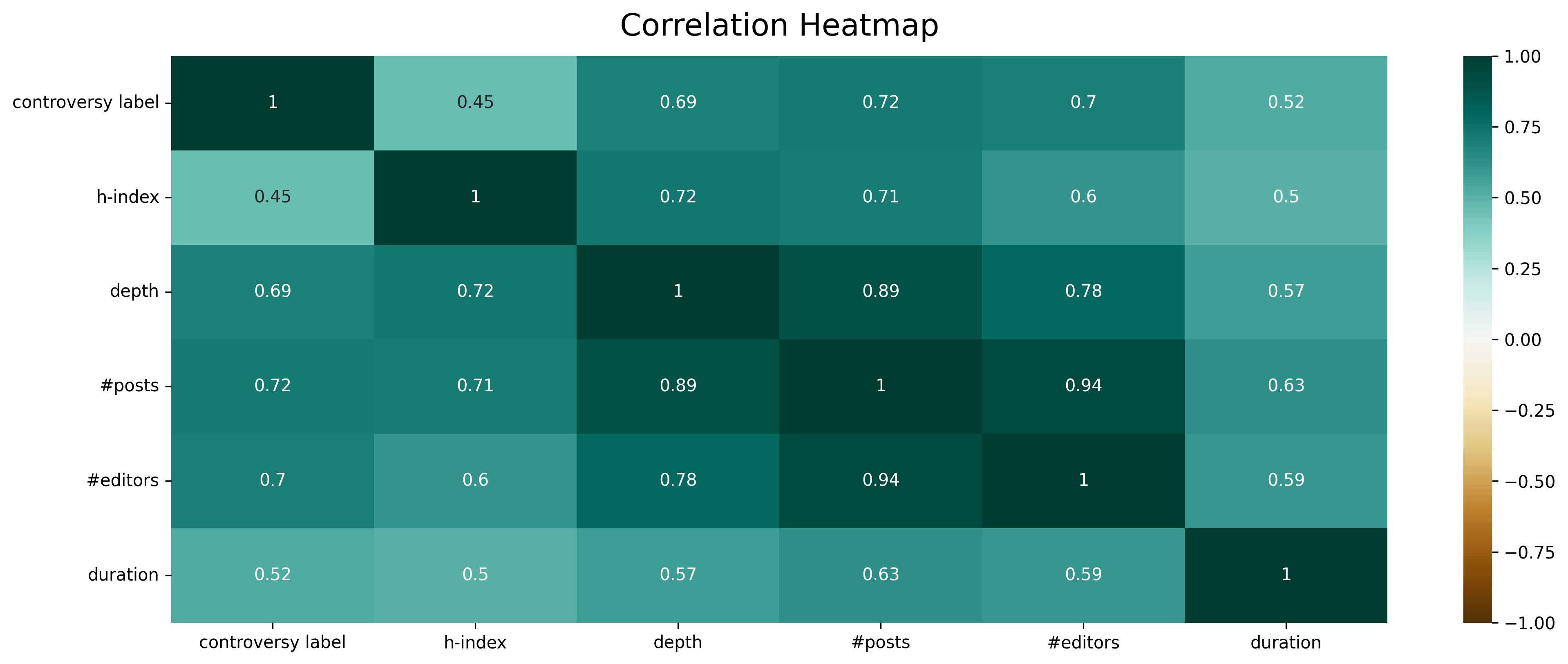}
\caption{Correlation matrix for features related to thread characteristics}
\label{fig:heatmap radial trees}
\end{figure}

Figure~\ref{fig:heatmap radial trees} presents the correlation matrix of these measurements. The results indicate that a controversy is correlated with the number of editors ($0.7$) and posts ($0.72$), and the depth of the thread ($0.69$) and less with the h-index ($0.45$).
It is worth noting that the positive correlation between controversies and the number of participants and posts is similar to our initial notion regarding detecting controversies. This may imply that these features can be an indication of controversies, but as mentioned in Section~\ref{sec:resultsRQ1} and \ref{sec:RQ2}, the content of the discussion and the followed practices for each discussion channel still need to be considered in controversy detection.
In addition, the h-index and the depth of the thread are highly correlated to the number of posts (h-index $0.71$, and depth $0.89$) and editors (h-index $0.6$, and depth $0.78$). This was expected since the number of posts increases with increasing levels. 
Furthermore, there is a high correlation between the number of editors and posts ($0.94$). This indicates that in controversial threads, it is likely to find many participants exchanging fewer posts each than a few participants exchanging many posts.
Finally, the duration of the discussion did not present a correlation with the controversy label ($0.52$) and presented a slight correlation with the other features (h-index $0.5$, depth $0.57$, number of posts $0.63$, and number of editors $0.59$), indicating that the time response has a small influence on the course of the discussion (see also Figure~\ref{fig:heatmap radial trees}). This may come from the editors' habits of editing Wikidata, which can vary in days or weeks. Therefore, the frequency of participation in discussions can vary, suggesting that every issue can have an unpredictable duration.

While depth seemed like a good characteristic for controversial threads, it might be biased due to the structure templates that editors use to raise an issue. As described in Section~\ref{sec:community and contributions}, editors can use specific templates to raise an issue, explain their proposal, give examples, and specify further information depending on the channel. Specifically, for \textit{Requests for comment} channel, the issues are generally related to the project's practices. For the \textit{Request for comment} we found that using the template editors describe the main issue at the beginning of the request and create sub-issues with suggested solutions or ideas for the community to discuss. This creates multiple sub-discussions under the main issue and increases the depth of the thread.

\begin{figure*}[h]
\centering
\includegraphics[width=\linewidth]{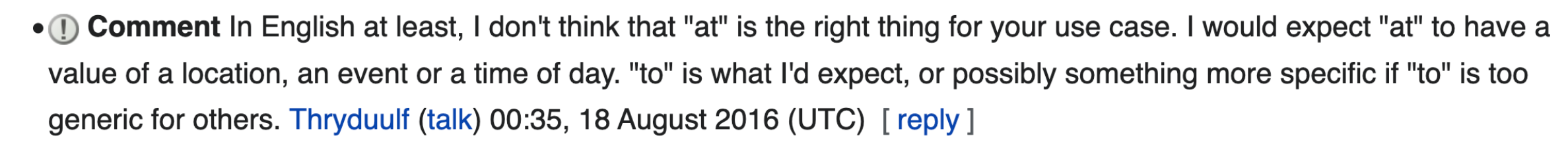}
\caption{An example for the combined use of code \textit{Challenge-Alternative} and \textit{Idea}.}
\label{fig:at}
\end{figure*}

\subsection{$RQ4$: How do editors argue when they disagree in Wikidata?}
\label{sec:RQ4}

\paragraph{Summary} To analyse how editors argue, we used the coding scheme from Table \ref{tab:ca}. In a few cases where the posts were long, including several arguments, we assigned more than one code. We found that editors most frequently disagree using a counter example ($33\%$). Counter example can be an example of a similar case or policies and rules.

Table \ref{tab:ca_results} shows the percentage of posts assigned with the argumentation scheme to the sample of $21$ threads that were identified as controversial in $RQ1$.
Results showed that editors frequently argue by disagreeing with the main issue or other opinions using \textit{Challenge - Counter example} ($33\%$), followed by \textit{Challenge - Alternative}, ($14\%$). For Wikidata, counter example could serve as an example of a similar case, but also as an example of practices and rules (see Figure~\ref{fig:find_a_grave} for counter examples). 
Consequently, the next most used argument was for agreement with other opinions. Editors use  \textit{Justification - Example}, $10\%$. Similar to the case of \textit{Challenge - Counter example}, \textit{Justification - Example} could serve as an example of similar cases, but also as an example of policies.

The codes \textit{Issue} and \textit{Idea} were the ones that pointed respectively to the main issues that were suggested for discussion in the three channels and any new ideas introduced during the discussion for taking a decision over the issue. 
Results showed that a new idea is introduced often ($10\%$). This code was usually assigned in combination with the opposing code \textit{Challenge-Alternative}. Figure~\ref{fig:at} shows an example for the use of both codes, where an editor disagrees with the proposal to create a property naming \textit{at}. 

\begin{table}[h]
\centering
\caption{This table presents the percentage of the codes assigned to the posts of 21 controversial discussions in order to analyse the type of arguments.}
\label{tab:ca_results}
\begin{tabular}{|l|l|}
\hline
\textbf{Code}      & \textbf{\%}                                                                                                                                   \\ \hline
Challenge - Counter Example & 33                                          \\ \hline
Challenge - Alternative     & 14                                     \\ \hline
Justification - Example     & 10                      \\ \hline
Justification - Evaluation  & 5                           \\ \hline
Issue                       & 3 \\\hline
Idea                        & 10 \\\hline
Elaboration                 & 9                                                                                       \\ \hline
Position - Agree            & 5    \\ \hline
Position - Disagree         & 8 \\ \hline
Decision - Under discussion & 0.1                                                                                       \\ \hline
Decision - Postponed        & 1                                     \\ \hline
Decision - Discarded        & 0.1                                     \\ \hline
Decision - Agreed           & 1                           \\ \hline
\end{tabular}

\end{table}

We used the \textit{Position} codes to find the agreements and disagreements in the threads for the cases where editors chose sides, mainly by voting, but did not express opinions to justify their choice. In this case, similar to codes for \textit{Challenge} where editors disagree with an opinion and \textit{Justification} where editors agree, editors more often \textit{Disagree} ($8\%$) than \textit{Agree} ($5\%$).

Finally, our coding scheme included codes relating to \textit{Decision} to capture the ratio of decision-making in controversial threads. We found cases where editors reached an \textit{agreement} ($38\%$) or the discussion was \textit{postponed} ($48\%$), and there were other different cases where the discussion was \textit{discarded} ($5\%$) or it is still \textit{open} ($9\%$). Figure~\ref{fig:find_a_grave} shows the status of this discussion at the top, which is ``consensus''.

\subsection{$RQ5$: What are the roles of editors in an argumentation?}
\label{sec:RQ5}
\paragraph{Summary} We found that the majority of participants are \textit{Rebels} ($46\%$), i.e., editors with strong ideas, followed by \textit{Loners} ($25\%$), i.e., editors who express their opinions with legitimate arguments, but do not engage in the discussion with more than two comments. In addition, we found a low frequency of \textit{Ignored Rebel} ($2.2\%$), suggesting that most of the opinions are taken into account, and a low frequency of \textit{Disruptors} ($1.3\%$), indicating low levels of vandalism.

We used the coding scheme from Table \ref{tab:actors} to assign one code for each editor. Table \ref{tab:actors_results} shows the results of the content analysis about the roles of editors in the argumentation.

All analysed threads included \textit{Rebels} and \textit{Loners}. We found that the majority of the posts ($46\%$) were posted by a \textit{Rebel}. This indicated that most frequently, editors had strong opinions and aimed to change others' opinions in their direction. We also found cases of controversy where a \textit{Rebel}, very early in a discussion, stated an opinion and then responded to every editor with an opposing opinion. In addition, many posts ($25\%$) were posted by a \textit{Loner}. These editors had minimal participation in the discussion but posted valid and useful arguments for the discussion. This often happened in the format of a voting process where editors voted and justified their opinion, but did not argue or interact further with other editors in the discussion.

The high number of posts by \textit{Rebels} ($46\%$), combined with the low number of posts by \textit{Leaders} ($8\%$), showed that editors did not often simply follow the ideas of others, but rather expressed and defended their own opinions. We also identified a low number of posts by \textit{Followers} ($8\%$), who did not necessarily change their opinions but often voted using other editors' opinions as justification.

We identified a low number of posts by \textit{Outsiders} ($9\%$). These editors participated in the voting process without justifying their votes or participating in the rest of the discussion. Furthermore, we found a low number of posts by \textit{Ignored Rebel} ($2\%$) or \textit{Disruptors} ($1\%$). This means that at the core of Wikidata, all editors who were willing to participate in a discussion received a response from the community. Additionally, there were no vandals among the editors. We did find one case of possible vandalism in the proposition of a \textit{Property for deletion}. In the early stages of this thread, there was some confusion when some editors suggested that the request may be fake based on the proposer's editing activity and the lack of description in the proposition. However, others expressed arguments as usual. Very soon, the disruption stopped, and the discussion continued.

Issues are usually started by a \textit{Loner} or a \textit{Rebel} in the \textit{Property Proposal} and \textit{Properties for Deletion} discussion channels. However, in \textit{Requests for Comments}, often, posts describing the main issue of the discussion did not include a signature in the end to specify the username. This made it difficult to identify the role of this editor in the thread.
In controversial discussions, \textit{Rebels} and \textit{Loners}, who are the most usual roles, most often used the argument \textit{Counter example}, with $39\%$ and $37\%$ respectively, followed by \textit{Alternative} with $17\%$ and $14\%$.

\begin{table}[t]
\centering
\caption{This table presents the percentage of the codes assigned to the posts of 21 controversial discussions in order to analyse the role of participants in argumentation.}
\label{tab:actors_results}
\begin{tabular}{|l|l|}
\hline
\textbf{Participant role}      & \textbf{\%}                                                                                                                                                                                                      \\ \hline
Rebel               &   46  \\ \hline
Loner    &                   25     \\ \hline
Leader              &   8                                                                                             \\ \hline
Follower            &  8\\ \hline
Outsider           &       9                                        \\ \hline
Ignored Rebel &                  2                                                                                \\ \hline
Disruptor           &           1                                                                                  \\ \hline
Others           &          1                                                                                                    \\ \hline
\end{tabular}
\end{table}

\section{Discussion}

In this work, we have studied Wikidata discussions to understand how editors disagree in collaborative KGs. We found disagreements, identified controversial threads, investigated the main themes, and explored their characteristics through a variety of measurements. Additionally, we studied the argumentation techniques and editors' roles in argumentation.

\textbf{$RQ1$} showed that $30\%$ of the analysed threads identified as controversial. Among the two channels related to properties, \textit{Properties for deletion} attract more disagreements and include longer discussions, than \textit{Property proposal}. In addition, in \textbf{$RQ2$} we found that for \textit{Property proposal} a few votes are enough to take decisions.  These results suggest that the community takes fast decisions on creating properties, and it is worth further analysis of whether these decisions impact the quality of Wikidata.

\textbf{$RQ2$} indicated that the most frequently identified theme of controversy in Wikidata is the process controversy ($52\%$).
This contrasts the results related to Wikipedia and open-source software development projects where the most frequent type is the task controversy \citep{arazy2013stay, filippova2016effects}. The difference in these results between the peer-production projects may come from the nature of the projects and the purpose of the analysed discussion channels. In collaborative KGs, the goal is to create a structural artifact using cognitive knowledge but also procedural rules, like editing item pages and following similar taxonomies with other cases in the project (similar to process controversy), while in Wikipedia, for example, writing an article is more related to cognitive knowledge (similar to task controversy). This particular difference in KGs compared to other peer-production projects may imply the need for clear policies or extra guidelines related to the process of editing and the hierarchy of data, i.e. classes, subclasses, and so on.

\textbf{$RQ3$} showed that measurements of radial trees, such as h-index, were not enough to indicate individually whether a discussion included controversies. Similar results can be seen in online discussions in Wikipedia \citep{laniado2011wikipedians} and Slashdot \citep{gomez2008statistical}. In Wikipedia, similar studies suggested that h-index is a good indicator for the depth of the discussion, but the chains of direct replies were a good indicator for contentious topics. Furthermore, in Slashdot, the authors found that many discussions shared the same h-index and suggested another measurement, the sum of h-index and the inverse of the number of posts, to break the ties \citep{gomez2008statistical}. However, a first indicator to identify controversies can be the size of the thread (number of posts), the number of participants in the discussion, and where the discussion is taking place (the discussion channel). Further investigation could reveal extra measurements for Wikidata.

In addition, \textbf{$RQ3$} suggests that the number of participants and posts presented a similar positive correlation with controversial threads ($0.7$ and $0.72$, respectively). In contrast, studies for the detection of conflicts in Wikipedia show number of posts present the highest correlation, proving that the most valuable feature in conflict prediction was the number of edits in the article talk pages, while the number of unique participants on the talk page is negatively correlated with conflict \citep{kittur2007he}. The results point to one fundamental difference between the two projects and highlight the significance of considering their different editing processes and their structures similar to other studies \citep{wikimediaContoversies,filippova2016effects,kittur2010beyond}.  

\textbf{$RQ4$} indicated that counter example using similar cases, and examples of policies and practices was the most used argument ($33\%$). This is similar to results in Wikipedia where the study of suggestions for the deletion of Wikipedia articles, using the Walton argumentation scheme, showed that among others \textit{rules} and \textit{evidence} comprise $36\%$ of the arguments \citep{schneider2013arguments}. These results highlight again the importance of clear and detailed policies. Our findings suggest that project designers and managers need to support members with easy-to-access and clear instructions regarding policies and guidelines in order to help them participate in discussions with legitimate and evident arguments.

Our argumentation results also identified that editors most often disagree when arguing (the sum of \textit{Challenge} codes was $47\%$), as compared to agreeing (the sum of \textit{Justification} codes was $15\%$). 
This contrasts with results from argumentation analysis in an open source software development project in which the authors found $13\%$ disagreement codes and $46\%$ agreement codes using data from GitHub using Toulmin's argumentation model \citep{wang2020argulens}. This difference may come from the difference in the datasets, where our analysis in Wikidata is focused on controversial threads, while the GitHub analysis chose five threads with a common topic based on the application used, heterogeneity and a high number of participants, and the variety of issues over this topic. Furthermore, we found differences between the two projects in the number of non-argumentative posts in the discussions. Looking at the \textit{Position} codes (\textit{agree} and \textit{disagree}, where discussion participants respond to an issue, idea, or argument, with a position which does not express argument), we found that in Wikidata, $25\%$ of codes were not argumentative, while in GitHub it was almost twice the size, i.e. $41\%$ of codes \citep{wang2020argulens}. The high number of posts with no argument in a long discussion with many participants may confuse the editors and make the final decision harder. In this case, a tool to summarise legitimate opinions and hide unrelated comments may support decision-making. Similar techniques were used to StackOverflow and helped with the flow of answers \citep{ren2019discovering}.

\textbf{$RQ5$} indicated that $8\%$ of editors participating in discussions were \textit{leaders} and another $8\%$ were \textit{followers}. This is similar to other OE projects like Schema.org \citep{kanza2018does}. Furthermore, the analysis revealed that a high number of participants, \textit{loners} ($25\%$), contributed with one or two legitimate posts but did not engage further in the discussion. The number of loners may indicate the reasons behind the lack of decision-making (we found $62\%$ of controversial discussions with no consensus). Participants who do not engage in discussions do not support their arguments or do not accept other valid arguments to conclude in decisions, thus leading the discussion to be closed because it remains inactive for a long, but with no consensus. These results, combined with our findings above that editors mostly disagree and that a high number of posts do not include arguments, may form explanations for the lack of decision-making.
All results indicate the need to change strategies in making decision when there is a long discussion with a high number of participants. A tool for flagging long discussions with controversies, and maybe a list of arguments included in the discussion, could support the community to make decisions.

The limitations of our study were the lack of information regarding participants' characteristics and the investigation of three specific discussion channels in Wikidata. Including information regarding the age of editors in Wikidata, their rights, and the rhythm of participation in editing activities could reveal more insights regarding how decisions are made in Wikidata and by whom. Furthermore, we analysed three discussion channels with topics related to the construction of the graph and suggestions regarding general policies and practices. Other discussion channels might include disagreements at smaller scale, but based on the topic of the channel the controversial themes might differ.

\section{Conclusion and Future work}

Community interactions, particularly disagreements, form the means to study how the community functions and makes decisions. In this work, we used descriptive statistics, thematic analysis, tree construction, statistical tests and content analysis to analyse disagreements in the collaborative KG Wikidata. 

We identified and explored three discussion channels as candidates that include disagreements, \textit{Request for comment, Properties for deletion} and \textit{Property proposal}. Our main findings were that quick decisions were taken in the creation of properties and that the majority of controversies ($52\%$) were related to policies and practices. Furthermore, more than half of the controversial discussions ($62\%$) did not lead to consensus. We found that a high number of editors ($25\%$) participated in discussions with one or two legitimate posts but did not engage further with the topic. At the same time, we revealed that Wikidata is an inclusive and peaceful community with all opinions taken into account within discussions and a very low rate ($1\%$) of vandalism. 

The findings presented in this study can be utilised to develop or enhance practices and tools that improve collaboration and communication within the Wikidata and other collaborative KE communities at large. Particularly, to enhance discussions including disagreements, collaborative KE communities could incorporate the following suggestions in their practices:
\begin{itemize}
    \item Providing clear and easily accessible information regarding community guidelines and policies is essential for enabling consistent editing behaviour and maintaining shared norms. Additionally, offering documentation on ontology evolution, such as class hierarchy and property usage, helps the editors to align their work with current practices and reduces ambiguity in data modelling decisions. Our findings indicated that the majority of arguments in discussions rely on these guidelines, policies, and practices. Therefore, while clear and easily accessible information enhances collaboration, it also has the potential to support disagreements and improve communication. Clear and accessible information could: improve the discussion flow; easier the process to reach consensus; make decisions which align with the principles of the project; and make decisions which align with the modelling practices.   
    \item Developing a tool to identify controversial discussions could draw attention to the community to provide support for consensus-building. Our findings suggest that key metrics -- specifically, the size of a discussion thread, the number of participants engaged, and the specific channel in which the disagreement occurs -- serve as effective initial indicators of controversy. 
    To enhance the performance of this tool, a second layer employing advanced technology such as large language models could be implemented. This approach could enable finer filtering, looking at the language and sentiment of the discussions. This dual-layered tool could improve the performance and accuracy of controversy detection, ultimately contributing to more constructive and informed discussions. 
    \item Creating a tool to summarise opinions and arguments in lengthy discussions. A tool to support decision-making in lengthy discussions could have several benefits for the community. These include increasing the rate of consensus, providing a summary archive for decisions, and motivating members to engage more in discussions. This could be feasible today with the advent of generative AI tools. Many writing software tools, like Grammarly\footnote{\url{https://app.grammarly.com/}} incorporate text summarisation and lists of main points in long documents. Furthermore, meeting summarisers like ScreenApp\footnote{\url{https://screenapp.io/features/meeting-summarizer}} transcribe and then summarise key topics in discussions. In online communities like Reddit\footnote{\url{https://redditinc.com/}} -- a news aggregation and social media forum platform -- contributors have already investigated how to summarise long conversations using the ChatGPT conversational agent.\footnote{\url{https://www.reddit.com/r/ChatGPT/comments/17qnjuv/any_plugin_that_can_summarize_reddit_posts/}} Therefore, a plugin for collaborative KE projects supporting discussions with a list of arguments could help the course of discussions and decision-making. 
\end{itemize}

Furthermore, our findings suggest several promising directions for future research. 
It would be worthwhile to investigate how quickly decisions, particularly those related to the creation of core entities like properties, affect the quality of the KG over time. Moreover, it would be interesting to study what is the rate of property creation and how this shapes the structure of the KG. 
In this vein, an essential future direction of this study would be to explore the connection of discussions and disagreements with the structure and quality of the KG.
Investigating the downstream effects of discussions on the KG evolution would provide valuable insights into the socio-technical dynamics of collaborative KE.
Future work could build on our findings to develop methodologies to trace the influence of discussions on editing-level decisions, thereby contributing to a deeper understanding of how human disagreements shape machine-readable knowledge. Additionally, another direction could extend this analysis to focus on editors and understand their interaction in collaboration. Specifically, it would be interesting to explore questions such as: \textit{Who participates more in discussions, newcomers or experienced editors? How do newcomers and experienced editors argue? Who most often closes discussions? Are there cases where closed issues reopen for discussion, and if yes, who most often reopens those issues?}


 
 
 
 
\section{Funding}
This project has received funding from the European Union’s Horizon 2020 research and innovation program under the Marie Skłodowska-Curie grant agreement no 812997 (CLEOPATRA ITN).

\section{Availability of data and materials} 
Our analysis and data are available on Github (\url{https://github.com/ElisavetK/Wikidata_disagreements})


\bibliographystyle{elsarticle-harv} 



\bibliography{sn-bibliography}

\end{document}